\documentclass[sigplan,nonacm]{acmart}

\usepackage{xkeyval}
\usepackage{hyperref}
\usepackage{hyperxmp}
\usepackage{gensymb}
\usepackage[normalem]{ulem}
\usepackage{xcolor}
\usepackage{amsmath,amsfonts}
\usepackage{algorithmic}
\usepackage{graphicx}
\usepackage{textcomp}
\usepackage{enumitem}
\usepackage{listings}
\usepackage{subcaption}
\usepackage{soul}
\usepackage{comment}
\usepackage{tikz}
\usepackage{xspace}
\usepackage{array}
\hypersetup{
    unicode=true,
    bookmarksnumbered=true,
    colorlinks=true,
    linkcolor={red!70!black},
    citecolor={red!70!black},
    urlcolor={blue!70!black},
    pdfborder={0 0 0}
}
\usepackage[capitalize,nameinlink]{cleveref}
\usepackage{flushend}
\usepackage{pifont}
\usepackage[table]{xcolor}

\newcommand{\myparagraph}[1]{\vspace{\smallskipamount}\noindent\textbf{#1.\xspace}}

\newcommand{\myparagraphemph}[1]{\vspace{\smallskipamount}\noindent\emph{#1.\xspace}}

\newcommand{\eg}{\emph{e.g.}\xspace}

\newcommand{\ie}{\emph{i.e.}\xspace}

\newcommand*{\rom}[1]{\uppercase\expandafter{\romannumeral #1\relax}}
\newcommand{\system}{StreamWise}
\newcommand{\app}{StreamCast}

\newcommand{\vs}{\emph{vs.}\xspace}

\newcounter{insightcounter}

\definecolor{darkgreen}{RGB}{0,150,0}

\begin{document}

\title[\system{}: Serving Multi-Modal Generation in Real-Time at Scale]{\system{}: Serving Multi-Modal Generation in Real-Time at Scale\vspace{15pt}}

\author{Haoran Qiu}
\affiliation{%
    \institution{Microsoft Azure Research}
    \city{Redmond}
    \country{USA}
}

\author{Gohar Irfan Chaudhry}
\authornote{Work done while interning at Microsoft Azure Research.}
\affiliation{%
    \institution{MIT CSAIL}
    \city{Cambridge}
    \country{USA}
}

\author{Chaojie Zhang}
\affiliation{%
    \institution{Microsoft Azure Research}
    \city{Redmond}
    \country{USA}
}

\author{Íñigo Goiri}
\affiliation{%
    \institution{Microsoft Azure Research}
    \city{Redmond}
    \country{USA}
}

\author{Esha Choukse}
\affiliation{%
    \institution{Microsoft Azure Research}
    \city{Redmond}
    \country{USA}
}

\author{Rodrigo Fonseca}
\affiliation{%
    \institution{Microsoft Azure Research}
    \city{Redmond}
    \country{USA}
}

\author{Ricardo Bianchini}
\affiliation{%
    \institution{Microsoft Azure Research}
    \city{Redmond}
    \country{USA\vspace{10pt}}
}

\date{}

\renewcommand{\shortauthors}{Qiu et al.}

\begin{abstract}
Advances in multi-modal generative models are enabling new applications, from storytelling to automated media synthesis.
Most current workloads generate simple outputs (\eg{}, image generation from a prompt) in batch mode, often requiring several seconds even for basic results.
Serving real-time multi-modal workflows at scale is costly and complex, requiring efficient coordination of diverse models (each with unique resource needs) across language, audio, image, and video, all under strict latency and resource constraints.

We tackle these challenges through the lens of real-time podcast video generation, integrating LLMs, text-to-speech, and video-audio generation.
To meet tight SLOs, we design an adaptive, modular serving system, \system{}, that dynamically manages quality (\eg{}, resolution, sharpness), model/content parallelism, and resource-aware scheduling.
We leverage heterogeneous hardware to maximize responsiveness and efficiency.
For example, the system can lower video resolution and allocate more resources to early scenes.

We quantify the trade-offs between latency, cost, and quality.
The cheapest setup generates a 10-minute podcast video on A100 GPUs in 1.4 hours (8.4\texttimes{} slower than the real-time) for less than \$25.
\system{} enables high-quality real-time streaming with a sub-second startup delay under \$45.
\end{abstract}

\maketitle

\section{Introduction}
\label{sec:intro}

\myparagraph{Motivation}
Advances in large language models (LLMs) are transforming a wide range of text-generation applications, from developer tools to education~\cite{githubcopilot,claudecode2025,xu2024llm_education}.
Building on this momentum, multi-modal content generation (audio, image, and video) is becoming more popular~\cite{flux2024,veo3,gao2025seedance,sora,wang2025wan,kong2024hunyuanvideo,kokoro}.
The next frontier is to support user-facing, real-time streaming applications at scale, enabling new experiences such as dynamic storytelling, personalized tutoring, and automated media creation~\cite{wang2025wan,kodaira2025streamdit}.

Although recent systems research has substantially improved the efficiency and scheduling of text-based LLM inference~\cite{jiang2023hexgen,patel2023splitwise,agrawal2024taming}, these techniques do not address the unique challenges posed by multi-modal, real-time generation.
Serving and orchestrating heterogeneous multi-modal models under strict end-to-end latency constraints remains expensive and introduces fundamentally new systems challenges.

Existing pipelines are typically batch workflows with high latency~\cite{flux2024,veo3,gao2025seedance,sora,wang2025wan,kong2024hunyuanvideo}.
Commercial systems remain expensive (over \$2 per minute) and cap outputs <10 seconds~\cite{veo3,sora} due to prohibitive computational expenses.
In contrast, streaming multi-modal generation demands low-latency pipelines that coordinate expensive AI models across modalities while meeting strict service-level objectives (SLOs).

These demands place significant strain on both hardware and software infrastructure.
Coordinating these diverse multi-modal models in real time (each with unique resource profiles)
while scaling remains an open systems challenge.

\myparagraph{Our work}
We study the systems challenges of real-time multi-modal generation through the lens of a representative application: podcast video generation.
Given an input (\eg{}, a research paper or news topic), the workflow produces a video of multiple characters engaging in a conversation about the topic.
This requires orchestrating LLMs, text-to-speech synthesis, image and video generation, and video-audio synchronization into a unified, low-latency pipeline.

To support this workload, we design and implement a modular, adaptive serving stack that balances latency, cost, and quality called \system{}.
The system supports streaming (real-time playback) with tunable fidelity to meet varying SLOs and budgets.
Our approach also accounts for compute heterogeneity.
It enables multi-level parallelism, content reuse, and quality adaptation.
For example, by rendering early scenes at lower resolution to reduce startup delay while maintaining high quality for later segments.

We characterize performance and resource trade-offs, quantifying how system-level decisions impact latency, cost, and quality.
For example, the smallest configuration generates a 10-minute podcast video on a single 8$\times$A100 GPU server in 3.7 hours;
insufficient for real-time needs.
\system{} achieves sub-second latency by combining A100 and H100 GPUs, at under \$40 per video when serving at scale.

\myparagraph{Summary}
We make the following main contributions:
\begin{itemize}[leftmargin=*]
\item We design a real-time multi-modal generation application and analyze its systems implications.
\item We characterize the trade-offs between latency, cost, and output quality, identifying key bottlenecks.
\item We identify and articulate systems-level optimization opportunities to further improve efficiency and scalability.
\item We build a modular serving system that enables parallel execution and quality-aware scheduling.
\item We quantify the cost of serving multi-modal generation workflows under real-time constraints.
\end{itemize}

\section{Podcast video generation}
\label{sec:podcastvideogen}

\begin{figure}
    \centering
    \includegraphics[width=0.95\linewidth]{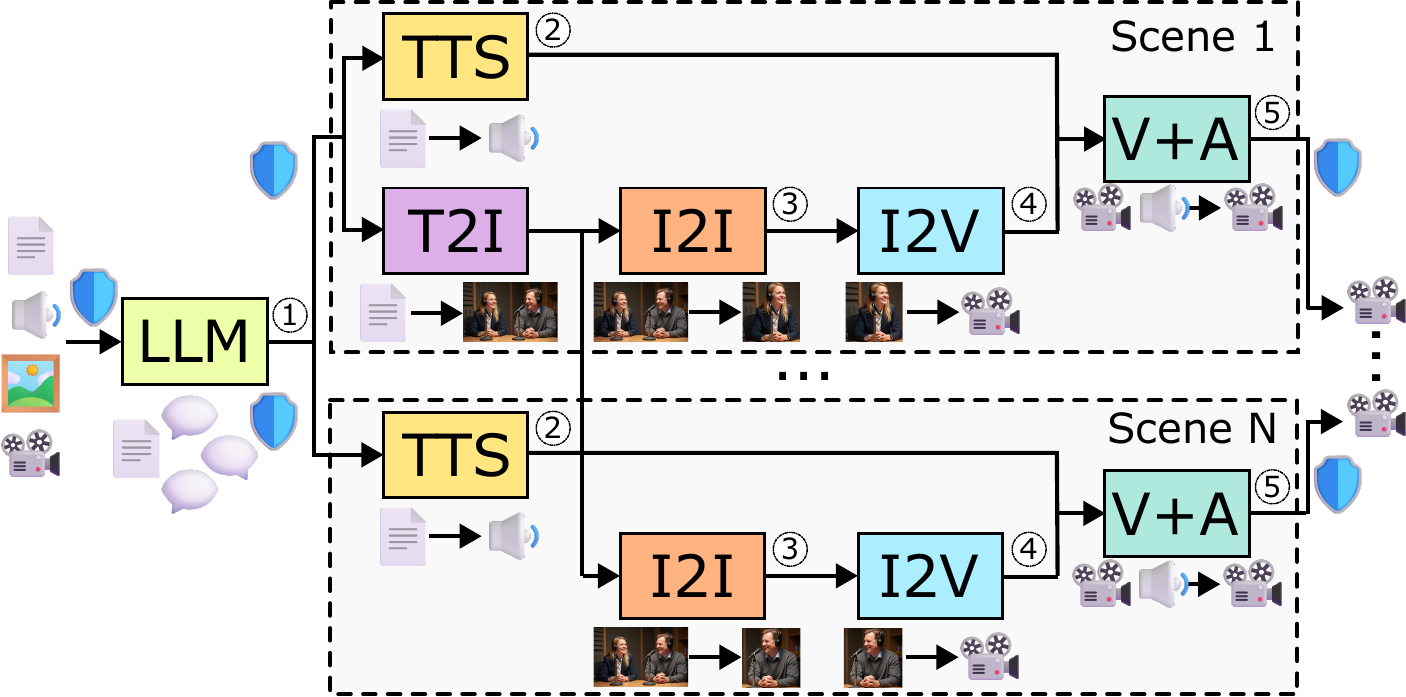}
    \caption{Workflow for video podcast generation.}
    \label{fig:workflow_podcast}
\end{figure}

\myparagraph{Overview}
We address the systems challenges of serving AI-generated multi-modal content in real-time using podcast video creation as a representative use case (extending beyond podcast audio~\cite{notebooklm}).
This workflow spans all modalities and involves diverse generative models operating under strict latency constraints.
Given a multi-modal input (\eg{}, a research paper with figures or a movie), it generates a podcast-style video with multiple characters discussing the input content.
\Cref{fig:workflow_podcast} shows the workflow:
(1) understand the input and generate a screenplay,
(2) generate audio,
(3) generate images,
(4) generate video, and
(5) synchronize video and audio.

\subsection{Workflow components}

\myparagraph{Screenplay generation}
From the input content, a multi-modal input \emph{LLM} generates the screenplay for the video.
We use prompting with few-shot examples (rather than fine-tuning or training) to produce scene-level descriptions, characters, and dialogues.
The screenplay describes each scene with the characters speaking and the sequence of shots.
For example, one shot with both characters, and one shot for each.
For every shot, the LLM generates a transcript of the spoken text.
As scenes and transcripts are generated, they trigger downstream stages.

\myparagraph{Image generation}
Scene and character descriptions are generated as prompts for image generation~\cite{ramesh2022clip} (\emph{T2I}).
The workflow generates consistent wide shots and close-ups as base images using multiple models and prompts (\emph{I2I}), ensuring coherence in style, design, and setting across shots.
This can also include static visual assets such as slides, text panels, or charts from the original input content.

\myparagraph{Audio generation}
The dialogue is synthesized with text-to-speech (\emph{TTS}) models using distinct voice profiles.
We apply emotion and prosody controls when supported.
We use specialized audio models to generate non-verbal sounds (\eg{}, ambient, effects, music).

\myparagraph{Video generation}
Using the base images as key frames, we generate shot-level video clips (\emph{I2V}).
For example, a 10-second shot may show one character speaking while others react, matching script timing and visuals.

\myparagraph{Video-audio sync}
For each shot, we align video and audio using lip synchronization and camera transitions (V+A).
Combining all shots across scenes, the final result is a cohesive podcast video that integrates all multi-modal assets.

\myparagraph{Safety}
AI-generated content poses safety, ethical, and legal risks.
To ensure responsible generation, our workflow has safeguards such as input validation, quality filtering, and post-generation compliance checks.

\subsection{Other workflows}
\label{sec:other_workflows}
Video podcast generation provides a versatile template and generalizes to many multi-modal applications.
\Cref{tab:other_workflows}
shows some examples with their inputs, multi-modal outputs, and compute characteristics.
These applications mainly involve modifying the LLM inputs (\ie{}, prompt design), adding stylistic modifiers (\eg{}, LoRAs~\cite{hu2021lora}), and tuning the data workflow.

For \emph{short} generation, the LLM supports multi-modal inputs (\ie{}, video understanding) and identifies key segments to reuse or regenerate (\eg{}, shorter scene).
For \emph{movie} generation, the LLM generates longer screenplays and videos with attention to narrative.
\emph{Lecture} and \emph{slide-persona} workflows use structured inputs (\eg{}, chapters or slide decks) and produce avatars synchronized with text-derived explanations and use static content (\eg{}, slides).
\emph{Dubbing} generates translated audio and uses video--audio sync.
\emph{Editing} skips most components and conditions the video-generation (\eg{}, cartoon style).
These examples show that our workflow extends naturally across a wide range of multi-modal applications.

\newcommand{\tablelow}{\cellcolor{green!20}Low}
\newcommand{\tablemedium}{\cellcolor{orange!20}Med}
\newcommand{\tablehigh}{\cellcolor{red!20}High}

\begin{table*}[t!]
\centering
\caption{Multi-modal generation workflows including example inputs, multi-modal outputs to generate, and characteristics.}
\label{tab:other_workflows}
\vspace{-10pt}
\footnotesize
\begin{tabular}{l lll cccc}
    \toprule
    Workflow &
    Example input &
    Generated output &
    Characteristic &
    Text &
    Audio &
    Image &
    Video
    \\
    \midrule
    \cellcolor{gray!20}Podcast~\cite{zhu2025paper2video} &
    \cellcolor{gray!20}Paper &
    \cellcolor{gray!20}Video podcast with characters discussing input &
    \cellcolor{gray!20}\emph{\Cref{sec:podcastvideogen}} &
    \tablelow &
    \tablelow &
    \tablelow &
    \tablehigh
    \\
    Short~\cite{barua2025lotus} &
    Movie &
    Extract key segments and generate highlight video &
    Heavy LLM &
    \tablehigh &
    \tablelow &
    -- &
    \tablelow
    \\
    Movie~\cite{meng2025holocine} &
    High-level plot &
    Multi-scene movie with characters, scenes,\ldots &
    Long output~\cite{imdb_database} &
    \tablelow &
    \tablelow &
    \tablemedium &
    \tablehigh
    \\
    Animated story~\cite{shi2025animaker} &
    Storyboard &
    Fully-animated story with characters and scenes &
    Style LoRA~\cite{hu2021lora} &
    \tablelow &
    \tablelow &
    \tablelow &
    \tablehigh
    \\
    Lecture~\cite{zhang2025transforming} &
    Textbook &
    Video with professor and supporting visuals &
    Static content &
    \tablelow &
    \tablelow &
    \tablelow &
    \tablemedium
    \\
    Slide persona~\cite{sun2022preavatar} &
    Slide deck &
    Embedded presenter persona narrating slides &
    Low resolution &
    \tablelow &
    \tablelow &
    \tablelow &
    \tablemedium
    \\
    Dubbing~\cite{zhang2024musetalk,waibel2023face} &
    TV show    &
    Translated and lip-synced preserving speaker &
    Adv. TTS~\cite{vibevoice} &
    \tablelow &
    \tablemedium &
    -- &
    \tablehigh
    \\
    Editing~\cite{sun2024diffusionedit,yu2025veggie} &
    Video &
    Modify components of the original video (\eg{}, style) &
    Heavy V2V &
    -- &
    -- &
    -- &
    \tablehigh
    \\
    Video chat~\cite{yan2024dialoguenerf} &
    Bot dialogue &
    Character delivering their side of the conversation &
    Short outputs &
    \tablelow &
    \tablelow &
    \tablelow &
    \tablemedium
    \\
    \bottomrule
\end{tabular}
\end{table*}

\subsection{Goals for multi-modal workflow serving}
Using our workflow to generate a 10-minute podcast (typical YouTube video length~\cite{Ferderick2025YouTubeSEO}) at 720p naively on an 8$\times$A100 GPU server takes 8.3~hours (50\texttimes{} slower than real time) and costs $\sim$\$70 in GPU time.
Over 99\% of the time and resources are consumed by video and video-audio generation.

\myparagraph{Real-time}
Most large-scale multi-modal generation systems run in batch mode, with users waiting minutes~\cite{sora,veo3}.
In contrast, we target real-time playback: video starts within seconds and streams uninterrupted.
This is captured by:
\begin{itemize}[leftmargin=*]
    \item \emph{Time to First Frame (TTFF)}:
    delay between input submission and display of the first frame.
    A low $TTFF$ improves perceived responsiveness.
    \item \emph{Time Between Frames (TBF)}:
    interval between generated frames.
    A steady $TBF$ ensures smooth video streaming.
\end{itemize}

To achieve real-time video generation at playback speed (one video second per second wall-clock), we require:
\[
\mathrm{TTFF}_{\mathrm{eff}}
= \max\!\left(
  \mathrm{TTFF},
  \overline{\mathrm{TBF}} \times \#\text{frames}
  - \text{video duration}
\right)
\]
For example, $\mathrm{TTFF}_{\mathrm{eff}}$ for a 10-minute video at 24 FPS and a TBF of 50 ms is 2 minutes, even if TTFF is 30 seconds.

\myparagraphemph{Deadlines}
This also imposes strict deadlines on when specific content must be ready.
For example, at 24 frames per second (FPS), frame 172 for scene 2 must be ready by 7.2~seconds;
with a TTFF of 1 second, the system must sustain a 36~ms TBF, relaxing to 42~ms once playback starts.

\myparagraphemph{Relaxed SLOs}
Some multi-modal generation requests come with more relaxed SLOs.
For example, a user may request that a video podcast be ready by 8\,AM.
Such cases have much looser deadlines, giving the scheduler greater flexibility and allowing shared components to be reused more effectively.

\myparagraph{Cost efficiency}
Large multi-modal models require costly hardware.
We try to minimize compute usage, maximize utilization, and account for datacenter constraints (\eg{}, power, thermal) that affect both provider and user costs~\cite{stojkovic2025tapas}.

\myparagraph{Output quality}
Users may tolerate minor quality degradation if content remains coherent and engaging.
The goal is to stay within an acceptable \emph{quality}.

\subsection{Monolithic integrated models \vs{} workflows}
\label{sec:monolith}

Alternatively, a single AI model could generate the entire video end-to-end, but it requires costly training and limits flexibility.
Instead, we adopt a modular design that reuses existing models, enabling easier orchestration, customization, and upgrades.
This allows us to quickly build new applications (\eg{}, short and dubbing).
Such workflows allow developers to upgrade components (\eg{}, TTS or video generation) without retraining and to rapidly adopt newer or more efficient submodels.
Similar modular architectures are widely used in production systems~\cite{kodaira2025streamdit,yuan2024mora}
and reflect a general trend towards modular agentic AI pipelines~\cite{chaudhry2025murakkab}.

Currently, there are no integrated models (open or closed) capable of handling complex, multi-stage long video generation (\eg{}, podcast-style video creation).
Existing models generate short clips (<25s SORA2~\cite{sora} and <8s Veo3.1~\cite{veo3}) from prompts, without coordinating dialogue, scenes, or modalities.
Even if such a monolithic model existed, it would still contain disaggregatable subcomponents.
For example, a diffusion-based one would have: text/image encoders, attention, diffusion, and video/audio decoders.
Similar to prefill/decode disaggregation~\cite{patel2023splitwise,zhong2024distserve} and chunking~\cite{agrawal2024taming} for LLMs, one would use such techniques to partition the video generation into shots (with their own deadlines).
This effectively transforms a monolithic model into a \emph{workflow} (similar to VAE-DiT disaggregation)~\cite{feng2025streamdiffusionv2} exposing scheduling and resource-allocation knobs without requiring architectural modularity.
While our insights target modular workflows, they also apply to monolithic models.
The main limitation is the reduced flexibility across specialized models (\eg{}, Flux~\cite{flux2024} \vs{} HiDream~\cite{hidream}).

\section{Multi-modal models characterization}
\label{sec:characterization}

To understand the system-level challenges of serving real-time multi-modal generation at scale, we analyze the key components of the podcast video generation workflow, with a focus on performance.
We focus on image and video generation, which together account for over 99\% of GPU time and dominate real-time serving costs.
We use Azure GPU VMs and unless otherwise stated, we use a full-node VM with 8 NVIDIA A100 GPUs~\cite{azure_a100}.

\subsection{Generative multi-modal models}
We evaluate a selection of pre-trained, open-source models from Hugging Face~\cite{hf_tasks}, using public leaderboards for image, video, and speech tasks~\cite{image_leaderboard,video_leaderboard,speech_leaderboard}.

We assess the model quality using Elo ratings~\cite{boubdir2024elo}.
This shows that open-source models (\eg{}, Wan~\cite{wang2025wan}, Hunyuan Video~\cite{kong2024hunyuanvideo}) are competitive with proprietary ones (\eg{}, Veo 3~\cite{veo3}, SeeDance~\cite{gao2025seedance}).
Based on available information~\cite{gao2025seedance}, close models typically use similar architectural foundations.

\Cref{tab:models} summarizes some of the evaluated models by category (\eg{}, image-to-video or I2V), architecture, and size (in FP16).
This shows that the field is evolving rapidly and over 10 new models have launched in the past three months.

\begin{table}[t!]
    \centering
    \caption{Generative models for podcast video generation.}
    \label{tab:models}
    \vspace{-10pt}
    \footnotesize
    \begin{tabular}{ccccc}
        \toprule
        Class   & Model & Architecture & Size & Release\\
        \midrule
        LLM     & Llama3.2~\cite{llama3}                & Transformer & \cellcolor{red!20}90B & 09/2024\\
                & Gemma3~\cite{gemma3}                  & Transformer & \cellcolor{red!20}27B & \cellcolor{green!20}03/2025\\
        A2T     & Whisper~\cite{radford2023whisper}     & Transformer & 1.5B & 11/2023 \\
        TTS     & Kokoro~\cite{kokoro}                  & Transformer & \cellcolor{green!20}82M & \cellcolor{green!20}01/2025 \\
                & XTTS~\cite{xtts}                      & Transformer & \cellcolor{green!20}400M & 09/2023 \\
                & OpenAudio~\cite{fish-speech-v1.4}     & Transformer & \cellcolor{green!20}500M & 11/2024 \\
                & Dia~\cite{diatts}                     & Transformer & 1.6B & \cellcolor{green!20}05/2025 \\
                & VibeVoice~\cite{vibevoice}            & Transformer & 7B & \cellcolor{green!20}09/2025 \\
        T2I     & SD3.5~\cite{sd3.5}                    & DiT & 8.1B & 03/2024 \\
                & Flux~\cite{flux2024}                  & DiT & \cellcolor{red!20}12B & 08/2024 \\
                & HiDream-I1~\cite{hidream}             & DiT & \cellcolor{red!20}17B & \cellcolor{green!20}04/2025 \\
                & Janus Pro~\cite{chen2025janus}        & Transformer & 7B & \cellcolor{green!20}01/2025 \\
        I2I     & YOLO~\cite{yolo11}                    & CNN & \cellcolor{green!20}68M   & \cellcolor{red!20}08/2016\\
                & Real-ESRGAN~\cite{wang2021realesrgan} & CNN & \cellcolor{green!20}16M   & 08/2021\\
                & FLUX Kontext~\cite{fluxkontext}       & DiT & \cellcolor{red!20}12B   & \cellcolor{green!20}07/2025 \\
                & Bagel~\cite{deng2025bagel}            & MoT & \cellcolor{red!20}14B   & \cellcolor{green!20}05/2025 \\
                & HiDream-E1~\cite{hidream}             & DiT & \cellcolor{red!20}17B   & \cellcolor{green!20}07/2025 \\
        I2V     & Wan 2.1~\cite{wang2025wan}            & DiT     & \cellcolor{red!20}14B & \cellcolor{green!20}02/2025\\
                & Wan 2.2~\cite{wang2025wan}            & DiT+MoE & \cellcolor{red!20}14B & \cellcolor{green!20}07/2025 \\
                & HunyuanVideo~\cite{kong2024hunyuanvideo}     & DiT & \cellcolor{red!20}13B & 12/2024 \\
                & FramePack~\cite{zhang2025framepack}          & DiT & \cellcolor{red!20}13B & \cellcolor{green!20}04/2025 \\
                & LTX-Video~\cite{HaCohen2024LTXVideo}           & DiT & \cellcolor{red!20}13B & 12/2024 \\
        V+A     & FantasyTalking~\cite{wang2025fantasytalking}   & DiT & 1.2B & \cellcolor{green!20}04/2025 \\
                & HunyuanAvatar~\cite{hu2025HunyuanVideo-Avatar} & DiT & \cellcolor{red!20}13B & \cellcolor{green!20}05/2025 \\
                & Sonic~\cite{ji2024sonic}                       & DiT & 1.1B & 12/2024 \\
        \bottomrule
    \end{tabular}
    \vspace{-4pt}
\end{table}

\myparagraph{LLMs}
These models typically use transformers~\cite{vaswani2017attention} and their performance characteristics have been well studied~\cite{patel2023polcapoweroversubscriptionllm,patel2023splitwise,vllm_benchmark}.
Inference involves a compute-intensive prefill phase that processes input tokens, followed by a memory-bound decoding phase that generates output tokens one at a time.
For example, running Gemma3 on a single A100 GPU can process over 7000 input tokens per second and generate one output token every 40 ms.
To scale across GPUs and servers, LLMs commonly use tensor and pipeline parallelism.

\myparagraph{TTS}
Modern text-to-speech models~\cite{hf_tts} also frequently adopt transformer-based architectures~\cite{vaswani2017attention}.
Typical models have a few hundred million parameters and require under 10 GB of memory, making them relatively lightweight.
Kokoro~\cite{kokoro}, a cost-efficient TTS model, generates one second of audio in under 1 ms on a single A100 GPU, with latency proportional to the number of input tokens (and roughly linear to output duration).
This can also run on CPUs which takes 500 ms to generate one second.
Larger models (\eg{}, Dia~\cite{diatts} and OpenAudio~\cite{fish-speech-v1.4}) have higher costs, with runtime increasing linearly with model size.
One can parallelize them across GPUs using sequence parallelism~\cite{li2021sequence}.

\begin{figure}
    \centering
    \includegraphics[width=1.00\linewidth]{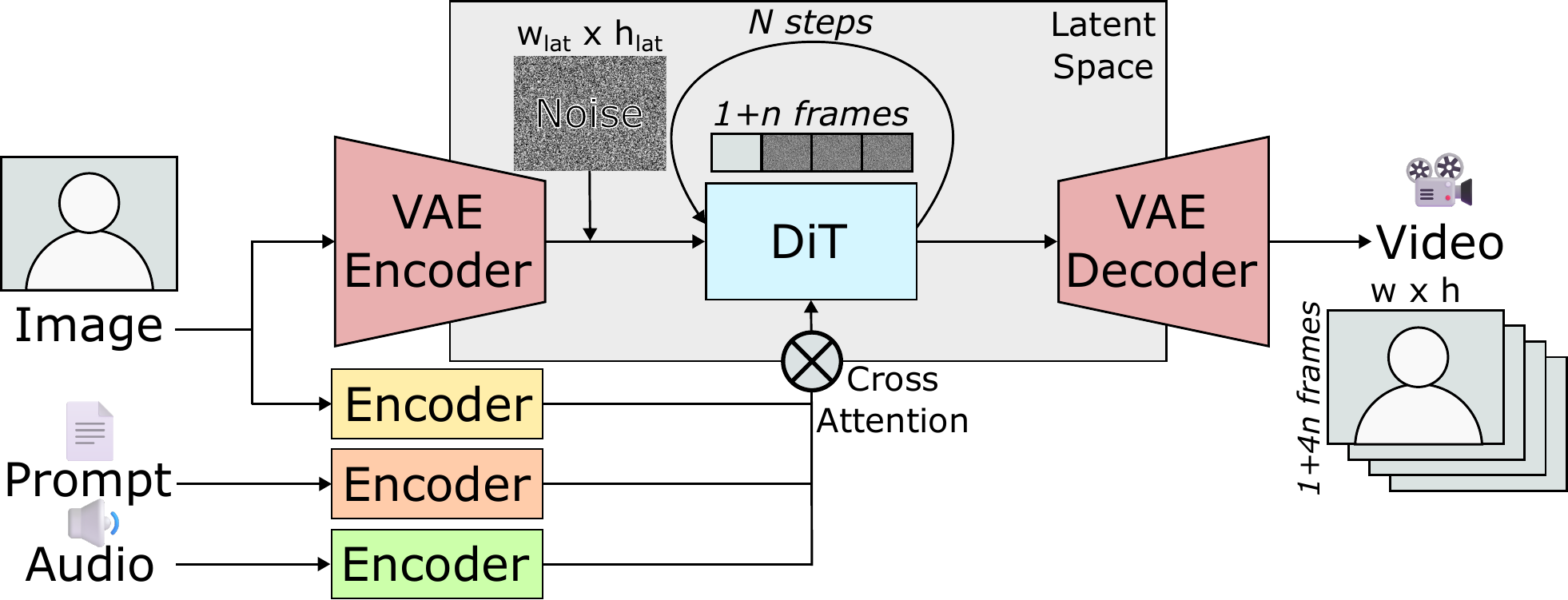}
    \caption{Video generation architecture using diffusion transformers and VAE with text, image, and audio inputs.}
    \label{fig:video_generation}
\end{figure}

\myparagraph{Image and video generation}
\Cref{fig:video_generation} shows the standard architecture for these models~\cite{hf_t2i,hf_t2v}, which are based on denoising diffusion~\cite{ho2020denoising}.
They often use Diffusion Transformer (DiT) architecture~\cite{saharia2022palette}, combining a Variational Autoencoder (VAE) with text/image encoders~\cite{kingma2013vae}.
Models like Wan 2.2~\cite{wang2025wan} adopt mixture-of-experts (MoE) DiT architectures.

The VAE encodes the starting frame into latent noise.
Most VAEs compress the input by a 8\texttimes{} in spatial resolution and expand the channels from 3 (RGB) to 16.
The input (text, image, or audio) is encoded and fed into the transformer via cross-attention.
DiT iteratively denoises the latent space until it resembles the target output.
These models typically use classifier-free guidance (CFG)~\cite{dhariwal2021diffusion}, running DiT with and without conditioning to balance fidelity and diversity.
Finally, the VAE decodes the latent into the final image/video.

\myparagraphemph{Text to image (T2I)}
Top-performing open-source models~\cite{image_leaderboard} such as FLUX~\cite{flux2024} and HiDream-I1~\cite{hidream} use DiT backbones (20--30 GB in FP16) to generate high-quality images from text prompts.
They typically use powerful text encoders like T5~\cite{raffel2020t5} or Llama3 and rely on dense attention mechanisms to capture fine-grained visual details.

\myparagraphemph{Image to image (I2I)}
DiT models encode an input image into tokens and transform them into latent space, conditioned on prompts or control signals.
They use common image encoders (\eg{}, CLIP ~\cite{radford2021clip}).
In this case, most VAEs compress the number of frames by 4 (leaving the first frame uncompressed) so for 1+80 frames, it compresses it into 21 latent frames.
Examples include InstantCharacter~\cite{tao2025instantcharacter}, FluxKontext~\cite{fluxkontext}, and HiDream-E1~\cite{hidream}.
Lightweight modules such as ControlNet~\cite{zhang2023controlnet,fluxupscaler} (\eg{}, using OpenPose or Canny edges) and IP-Adapter~\cite{ye2023ipadapter} inject external signals into the diffusion process.
Bagel~\cite{deng2025bagel}, for instance, uses a Mixture of Transformers (MoT) to perform structure-aware transformations.

For basic image generation, we can use cheaper models like YOLO~\cite{yolo11} which are based on Convolutional Neural Networks (CNNs).
These smaller models can recognize characters and generate zoomed-in images.

\myparagraphemph{Text/image to video (T2V and I2V)}
These models~\cite{hf_t2v} share the same architecture as image generation but de-noise multiple frames instead of a single image.
For example, Wan2.1~\cite{wang2025wan} can generate up to 1+80 frames (1 starting frame and 80 generated ones) at the same time which is 5.1 seconds at 16 frames per second.
Stable Video Diffusion~\cite{blattmann2023stable} and HunyuanVideo~\cite{kong2024hunyuanvideo} have similar implementations and are trained with a specific \emph{frame rate}.
For example, Wan uses 16 frames per seconds (FPS) while HunyuanVideo uses 30 FPS.

\myparagraphemph{Resolution}
Since computational cost increases with the number of pixels, directly generating high-definition content can be prohibitively expensive.
To address this, lightweight up-scaling models~\cite{wang2021realesrgan} can be applied to enhance resolution and temporal consistency.
LTX~\cite{HaCohen2024LTXVideo} adopts this strategy.%

\myparagraphemph{Video duration}
Due to memory constraints during both training and inference, most models restrict video generation to short clips (typically <5 seconds).
A common workaround is to generate the maximum allowed frames, then feed a few frames from the end as input for the next segment.
However, this can break temporal continuity and lead to visual drifting.
FramePack~\cite{zhang2025framepack}, built on HunyuanVideo~\cite{kong2024hunyuanvideo}, addresses this by compressing less important frames in latent space, enabling attention over the entire video.

\myparagraphemph{Other model architectures}
There are models based on a traditional transformer architecture like Janus~\cite{chen2025janus} and LlamaGen~\cite{sun2024llamagen}.
These models generate one token at a time that represents a patch of the image.
Currently, these models are neither popular~\cite{hf_tasks} nor fully mature;
their output quality remains lower~\cite{image_leaderboard}, while performance is comparable.

\myparagraph{Video and audio synchronization}
Models like Sonic~\cite{ji2024sonic} and RealTalk~\cite{ji2024realtalk} are specifically trained for video-audio synchronization.
Other models add a cross-attention layer to extend regular video generation models and align with the audio.
For example, FantasyTalking~\cite{wang2025fantasytalking} extends Wan~\cite{wang2025wan} and HunyuanVideo-Avatar~\cite{hu2025HunyuanVideo-Avatar} extends HunyuanVideo~\cite{kong2024hunyuanvideo}.
Their inference time is similar to plain video generation models (they scale linearly to the length of the video, the number of de-noising steps, and the video resolution).

\subsection{Performance breakdown}
\myparagraph{Request serving latency}
We focus on diffusion-based image-to-video generation (\ie{}, Wan 2.1), as it is the most expensive and it encompasses other models.
We highlight the differences with other models throughout the discussion.

\begin{figure}
    \centering
    \includegraphics[width=\linewidth]{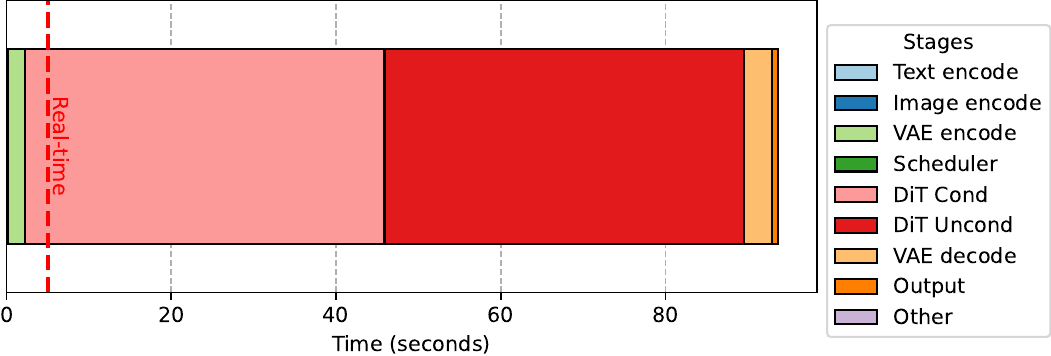}
    \includegraphics[width=\linewidth]{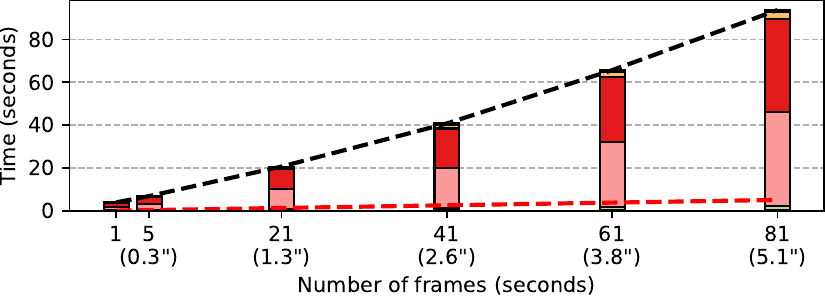}
    \includegraphics[width=\linewidth]{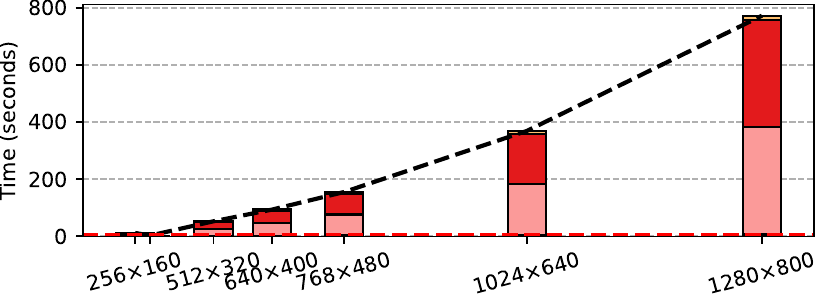}
    \includegraphics[width=\linewidth]{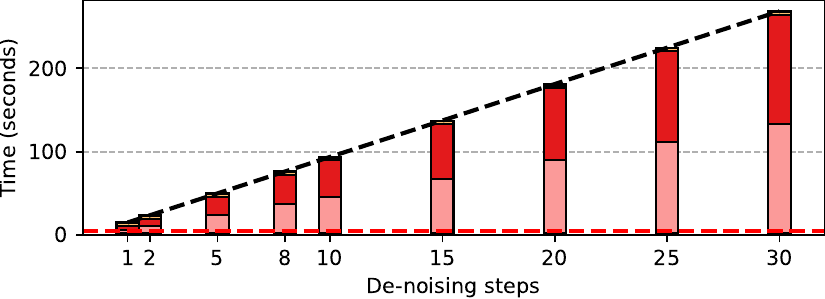}
    \includegraphics[width=\linewidth]{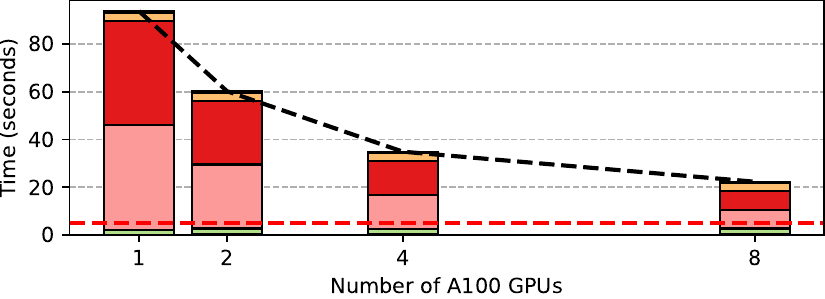}
    \caption{Latency to generate an 81-frame (5.1-second at 16 FPs) video at $640\times{}400$ (16:10 aspect ratio) using 10 diffusion steps on A100 GPUs.
    It includes the sensitivity to each of these parameters.
    The red dashed line indicates real time.
    }
    \label{fig:char_latency}
    \vspace{-10pt}
\end{figure}

\Cref{fig:char_latency} shows that generating a $\sim$5.1-second video (1 initial + 80 generated frames at 16 FPS) takes $\sim$93 seconds on one A100 GPU ($\sim$18 seconds per second generated video).
Most of this time is spent on diffusion (both conditional and unconditional DiT) and some on VAE encode/decode, while text and image encode are relatively lightweight.
Thus, text-to-video latency is similar to that of image-to-video.

For reference, commercial solutions such as Veo 3~\cite{veo3} and Sora~\cite{sora} take 2--4 minutes to generate 8 seconds for the cheaper plan and under 60 seconds for the more expensive one (7--30 seconds per second generated)~\cite{veo3costanalysis}.

Video-audio synchronization (\eg{}, Fantasy Talking) adds audio encoding and a cross-attention layer~\cite{qiu2025modserve} to the base model (\eg{}, Wan 2.1).
Audio encoding is even faster than text or image encoding, and the added cross-attention has negligible performance impact.

\myparagraph{Frames}
\Cref{fig:char_latency} shows that inference time is almost linear to the number of frames.
To generate 1 frame (62.5 ms) it takes $\sim$66 seconds/second,
for 1.3 seconds it takes $\sim$23, and for 5.1 seconds it takes $\sim$18 seconds/second.
Generating longer videos is slightly more efficient.
Latency for image generation is comparable to video generation for one frame.

\myparagraph{Resolution}
Because VAEs typically apply an 8:1 compression (\ie{}, one latent pixel per 8 input pixels), not all resolutions are allowed.
For example, 480p at a 16:9 aspect ratio ($854\times{}480$) has a width that is not divisible by 8, requiring padding or cropping to fit the model.
In contrast, resolutions with 16:10 or 5:4 aspect ratios more often align with the VAE 8\texttimes{} down-sampling, making them better suited.

\Cref{fig:char_latency} shows that latency increases with the number of pixels (width $\times$ height).
Increasing resolution from $640\times400$ to $1280\times800$ results in 4\texttimes{} more pixels and approximately 4\texttimes{} higher latency.
Note that higher resolutions need more de-noising steps to achieve similar quality.

\myparagraph{Steps}
\Cref{fig:char_latency} shows DiT latency increases linearly with the denoising steps.
More steps lead to higher output quality.
With only 1--2 steps, the image quality is poor, but objects become distinguishable around 5 steps.
Most images are sharp by 10 steps, and reaches peak quality at 20--30 steps.

\myparagraph{Number of GPUs}
To reduce latency, we can execute these models across multiple GPUs.
A commonly used method for parallelizing DiT-based models is Unified Sequence Parallelism (USP)~\cite{fang2024unified}, which combines Ulysses~\cite{jacobs2023deepspeed} and Ring attention~\cite{liu2023ring} to distribute the latent space across GPUs.
If the model employs CFG, we can further parallelize the conditioned and unconditioned DiT passes across GPUs.

\Cref{fig:char_latency} shows that USP across multiple GPUs on a single server significantly reduces the DiT execution time.
Note that USP applies only to the DiT stage and the VAE stages remain unchanged.
Using 8 GPUs, we achieve over a 5\texttimes{} reduction in DiT time compared to single-GPU execution.

\myparagraph{Batching}
To increase efficiency, one can batch and process multiple requests together~\cite{vllm,yu2022orca}.
This works well for image and text encoding, which scale almost perfectly.
However, DiT is already near compute saturation, and VAE is fully saturated, leaving little headroom.
Batching four requests improves DiT efficiency by less than 5\%, and for VAE, batching increases runtime due to execution overhead.

\myparagraph{Model loading time}
Serving large models at scale requires accounting for GPU loading time.
For example, Wan, with 14B parameters, has an image size of $\sim$90~GB. 
Loading its FP16 weights into GPU takes $\sim$30 seconds, and the first warm-up request (triggering compilation) requires $\sim$80 seconds.
Once loaded, the model occupies $\sim$48~GB of GPU memory.

Loading time scales with model size.
For example, Flux (12B parameters) loads in $\sim$10 seconds.
However, it needs $\sim$3 minutes to warm-up and uses $\sim$33~GB of GPU memory.

\subsection{Impact of hardware}
The performance of these models is highly dependent on the underlying hardware.
We benchmark them across four GPU generations and a CPU for reference (\Cref{tab:hardware}).

\begin{table}[t!]
    \centering
    \caption{Hardware, price per hour, and power.}
    \label{tab:hardware}
    \vspace{-10pt}
    \footnotesize
    \begin{tabular}{ccccrrr}
        \toprule
        Type & Model & Year & \#GPUs & Reserved & Spot & TDP\\
        \midrule
        CPU & Intel EMR & 2024 & - & \cellcolor{green!20} \$2.33  &  \cellcolor{green!20} \$0.83 & \cellcolor{green!20} 350 W\\ %
        GPU & V100~\cite{nvidia_v100} & 2017 &8 & \$10.79 &  \$3.97 & 8$\times{}$300 W\\ %
        GPU & A100~\cite{nvidia_ampere} & 2020 & 8 & \$14.42 &  \$8.52 & 8$\times{}$400 W\\ %
        GPU & H100~\cite{nvidia_hopper} & 2022 & 8 & \cellcolor{red!20} \$43.16 & \cellcolor{red!20} \$32.22 & \cellcolor{red!20} 8$\times{}$700 W\\ %
        GPU & H200~\cite{nvidia_hopper} & 2024 & 8 & \cellcolor{red!20} \$45.22 & \cellcolor{red!20} \$33.76 & \cellcolor{red!20} 8$\times{}$700 W\\ %
        GPU & GB200~\cite{nvidia_gb200} & 2025 & 4 & \cellcolor{red!20} \$57.67 & \cellcolor{red!20} \$43.04 & \cellcolor{red!20} 4$\times{}$1200 W  \\
        \bottomrule
    \end{tabular}
\end{table}

\myparagraphemph{Type and generation}
We focus on NVIDIA GPUs due to their broader availability and mature software ecosystem.  
While alternative accelerators (\eg{}, AMD MI300 or Google TPUs) can theoretically support these models, practical adoption remains limited due to poor software support.
On the other hand, CPUs are too slow to run medium-sized models efficiently.
They can handle smaller, cheaper models like Kokoro or YOLO, but with a slowdown of up to 60\texttimes{}.

\myparagraphemph{Availability}
GPUs are in high demand and subject to regional constraints.
Across 60 major regions of a leading cloud provider, only 1 offered V100, while A100, H100, and H200 were available in 18, 14, and 5 regions respectively and GB200 is widely unavailable at this time.
Even in these regions, quotas are tightly constrained.
Due to these constraints, we later explore mixing GPU types from multiple regions.

\myparagraphemph{Features}
Most DiT-based models (including Flux and Wan) leverage FlashAttention~\cite{dao2022flashattention,dao2023flashattention2,shah2024flashattention3}, a family of attention kernels optimized for NVIDIA Ampere~\cite{nvidia_ampere} and newer GPUs.
These optimizations deliver over 20\texttimes{} performance improvements but are incompatible with V100-class hardware~\cite{nvidia_v100}.
FlashAttention-3, in particular, is tailored for Hopper and newer architectures~\cite{shah2024flashattention3}, providing additional speedups.

\myparagraphemph{Quantization}
Modern GPUs include specialized compute units for data types beyond FP32, such as FP16, FP8 (Hopper+~\cite{nvidia_hopper}), and FP4 (Blackwell+~\cite{nvidia_gb200}).
Most open-source models use FP16 weights, though some unofficial implementations support FP8~\cite{fluxfp8}.
Quantization reduces GPU memory usage and improves latency~\cite{rokh2023quantization}.
For example, running Flux with FP8 reduces inference latency by 30--40\% on H100 GPUs.

\begin{figure}
    \centering
    \includegraphics[width=0.95\linewidth]{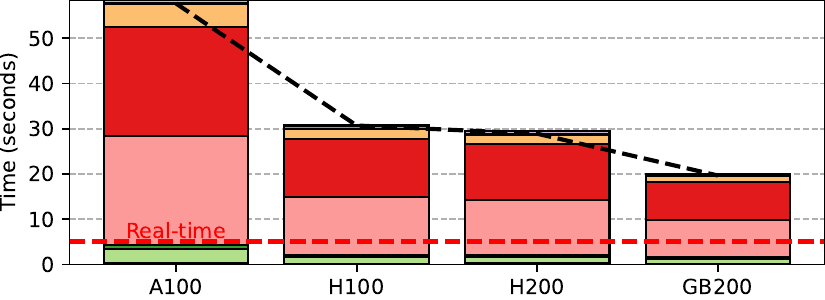}
    \caption{Latency sensitivity to the NVIDIA GPU generation with 4 GPUs in a single server (limited to match GB200).}
    \label{fig:gpu_type}
\end{figure}

\myparagraph{Latency}
\Cref{fig:gpu_type} shows that upgrading from A100 to H100 reduces the latency for Wan 2.1 by $\sim$1.9\texttimes{}.
H200, with increased memory bandwidth and capacity, improves performance by an additional 5\%.
GB200 further reduces latency compared to A100 by $\sim$2.9\texttimes{} (note that we do not use specific optimizations for Blackwell and use the default FP16).
These improvements benefit all stages, especially the most time-critical ones:
DiT denoising steps and VAE encoding/decoding.
It also reduces the model loading time.

\myparagraph{Cost}
\Cref{tab:hardware} includes the best GPU pricing for three major cloud providers:
EC2~\cite{ec2}, Azure~\cite{azure}, and GCP~\cite{gcp}, and validate it using a third-party aggregator~\cite{cloudprice}.
All prices correspond to the longest available reservation term (\eg{}, 3 years).
As a cheaper option, we include the Spot VM pricing.
Pricing varies significantly by region (often by >2\texttimes{}).

A100 GPUs are $\sim$60\% more cost-efficient than H100s and $\sim$2.7\texttimes{} than GB200.
This trade-off is important when targeting real-time serving at low cost.
We also need to factor in static costs such as model loading times and GPU idle times across the entire system.
For comparison, commercial video generation models like Veo 3~\cite{veo3} cost around \$0.20 for a maximum of 8 seconds at 24 FPS at 720p~\cite{veo3costanalysis}.

\myparagraph{Energy}
Relative to A100, H100 and H200 reduce total energy consumption by 7\% and 12\%, respectively, while GB200 consumes 2\% more energy despite being almost 3\texttimes{} faster.
Note that this only accounts for GPU and not for the rest of the system (CPU, memory, NVLink and others).
This highlights that performance gains alone do not directly translate into energy savings, particularly as power may not scale linearly across generations.
These trends motivate frequency scaling as a practical mechanism for improving energy efficiency without sacrificing real-time latency.

\myparagraph{Frequency scaling}
For completeness, we analyze the impact of GPU frequency scaling on A100, varying it from 200~MHz up to the maximum of 1.4~GHz.
Both the DiT and VAE stages show high sensitivity:
a 15\% frequency reduction increases runtime by 8\%, while a more aggressive 45\% cap results in a 52\% slowdown.

The A100 GPU has a thermal design power (TDP) of 400W.
Power consumption scales roughly quadratically with frequency, ranging from 63W when idle to 400W at peak.
Reducing frequency by 15\% lowers peak power by 23\%.
A 45\% reduction cuts it by 48\%.
At the highest frequency, average power remains within 10\% of the peak, indicating sustained high utilization with little idling.
The most energy-efficient range is between 800--1000MHz, reducing energy consumption by over 20\%.
Other GPU generations show similar trends when normalized to their TDP (\Cref{tab:hardware}).

GPU temperature closely tracks power usage and gradually stabilizes over time.
We observe inter-GPU thermal variation, with some GPUs consistently running hotter depending on their physical location and manufacturing~\cite{stojkovic2025tapas}.

\subsection{Distributed inference across multiple servers}
\begin{figure}
    \centering
    \includegraphics[width=\linewidth]{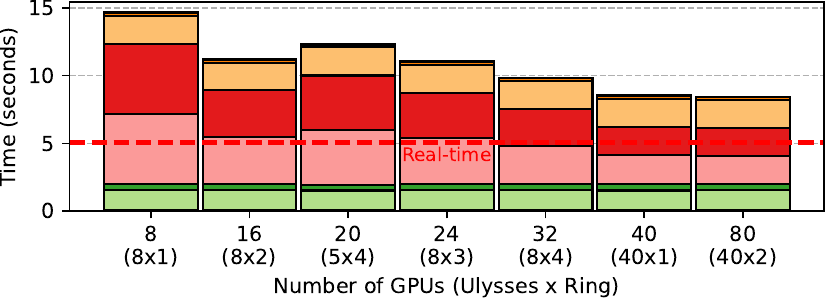}
    \vspace{-20pt}
    \caption{Sensitivity to the number of H200 GPUs across servers using USP with Ulysses and Ring attention.}
    \label{fig:char_parallel_multiple}
    \vspace{-10pt}
\end{figure}

Despite these improvements, latency is still below real-time.
We leverage InfiniBand~\cite{infiniband} connectivity to parallelize beyond a single server~\cite{patel2023splitwise}.
\Cref{fig:char_parallel_multiple} shows the latency going from a server with 8$\times$H200s to 10 servers with 80 GPUs.
Using five H200 servers (40 GPUs) enables real-time DiT performance when pipelining the VAE stages.
However, efficiency remains low:
achieving less than 18\texttimes{} latency speedup requires 40\texttimes{} more resources (\ie{}, from 1 to 40 H200 GPUs).

\myparagraph{Parallelism constraints}
USP is limited by both model architecture and input size.
For example, Ulysses~\cite{fang2024unified} parallelizes across attention heads (\eg{}, 40 in Wan), making it incompatible with setups like two 8$\times$GPU servers.
In addition, sequence length (linked to output resolution) must satisfy divisibility constraints to avoid padding or cropping.
It must be divisible by both the number of attention heads and GPUs.
Combined with VAE compression (\eg{}, 8\texttimes{}), these constraints narrow the set of effective 16:9 resolutions.
To improve scheduling flexibility, we adopt alternative aspect ratios such as 16:10 (8:5) and 5:4.

\section{Serving multi-modal generation}
\label{sec:system}

To serve our video podcast generation workflow in real time, we present potential system strategies and propose an end-to-end architecture that optimize each stage of multi-modal generation to balance latency, cost, and output quality.

\subsection{Systems opportunities for efficiency}

Building on the requirements of our workflow and the characteristics of its foundational multi-modal models, we outline system-level strategies to balance latency, cost, and quality.

\myparagraph{Reducing latency}
To generate multi-modal content in real-time, it is critical to minimize latency.

\myparagraphemph{Deadline-aware scheduling}
Strict deadlines (especially for early shots) require prioritization of latency-critical segments.
Deadline-aware scheduling allocates resources upfront for these segments while relaxing constraints for later ones, ensuring responsiveness.

\myparagraphemph{Parallelism}
To reduce latency, we can exploit multiple GPUs at several levels:
(1) independent \emph{stages} (\eg{}, audio and image generation) can run concurrently once previous ones (\eg{}, screenplay generation) finish;
(2) \emph{scenes and shots} can be generated independently, enabling spatial and temporal distribution across servers; and
(3) large \emph{models} support scale-out via sequence~\cite{narayanan2021efficient}, tensor~\cite{shoeybi2019megatron}, and pipeline parallelism~\cite{huang2019gpipe}.

\myparagraphemph{Disaggregation}
To enable pipelining and reduce latency, model components can be disaggregated~\cite{patel2023splitwise, zhong2024distserve}.
For example, separating DiT and VAE stages allows decoding to start in latent space while DiT is still denoising.
This overlap improves throughput and responsiveness.%

\myparagraph{Minimizing cost}
Our workflow introduces flexibility to reduce costs and improve resource efficiency.

\myparagraphemph{Hardware-aware}
Lightweight tasks may run on CPUs or older GPUs, while latency-critical stages use newer GPUs.
Components can scale independently, and newer hardware can be selected for time-sensitive segments.

\myparagraphemph{Fine-grained scaling}
Models can be provisioned independently based on workload demand.
In addition, disaggregation enables independent component scaling, offering precise control over resource allocation for latency-sensitive paths.
For example, we can provision 16 instances for I2V DiT, 2 for I2V VAE, and 1 for TTS.

\myparagraphemph{Spot and multi-region deployments}
To further reduce costs, we can aggressively use Spot resources, mitigating eviction risk via over-provisioning~\cite{yi2010reducing}.
To leverage available resources across multiple regions, we can aggregate them and orchestrate workflows accordingly~\cite{jiang2023hexgen,mei2025helix}, accounting for bandwidth and latency constraints.

\myparagraphemph{Multiplexing}
Serving multiple users concurrently allows better resource utilization and amortizing fixed costs, further improving overall system efficiency.
\emph{Batching} improves throughput but can increase latency~\cite{yu2022orca}.
Fine-tuned batching strategies can balance responsiveness and efficiency without overloading compute-intensive components.

\myparagraphemph{Frequency tuning}
Adjusting GPU frequencies allows scheduling to account for power, energy, and thermal limits.
This lowers infrastructure demands and reduces hosting costs.

\myparagraph{Quality}
To balance responsiveness, user experience, and cost, we can also tune output quality.

\myparagraphemph{Configuration selection}
Selecting appropriate models and configurations (\eg{}, de-noising steps, resolution) at each stage is crucial.
High-quality settings can be reserved for critical or visually prominent shots, while lighter configurations maintain real-time performance elsewhere.

\myparagraphemph{Adaptive quality} %
Early shots may use lower-fidelity models or reduced resolution to meet deadlines, with fidelity increasing as resources allow.
This enables real-time adjustment of script length, resolution, and model complexity based on system load and user expectations.

\myparagraphemph{Static content}
Pre-created intros, backgrounds, or interpolated visuals can mask delays, reduce load, and improve responsiveness.
During time-critical segments, static elements (\eg{}, slides, ads) absorb latency, while lightweight overlays enable low-cost personalization.

\subsection{\system{} overview}
\label{sec:system_overview}

We propose \system{}, an end-to-end solution for serving multi-modal content generation at scale.
Designed to support efficient real-time streaming, it addresses the unique demands of these workloads by building on the challenges and opportunities outlined in prior sections.

\system{} builds on well-known techniques (\eg{}, DAG scheduling, disaggregation) but its novelty lies in adapting and orchestrating them for real-time, diffusion-based multi-modal generation workflows.
This workload is more complex, compute-intensive, and latency-critical than prior LLM serving optimizations which rely on a two-stage workflow with prefill and decode~\cite{patel2023splitwise,zhong2024distserve,mei2025helix,jiang2023hexgen}).

\begin{figure}
    \centering
    \includegraphics[width=\linewidth]{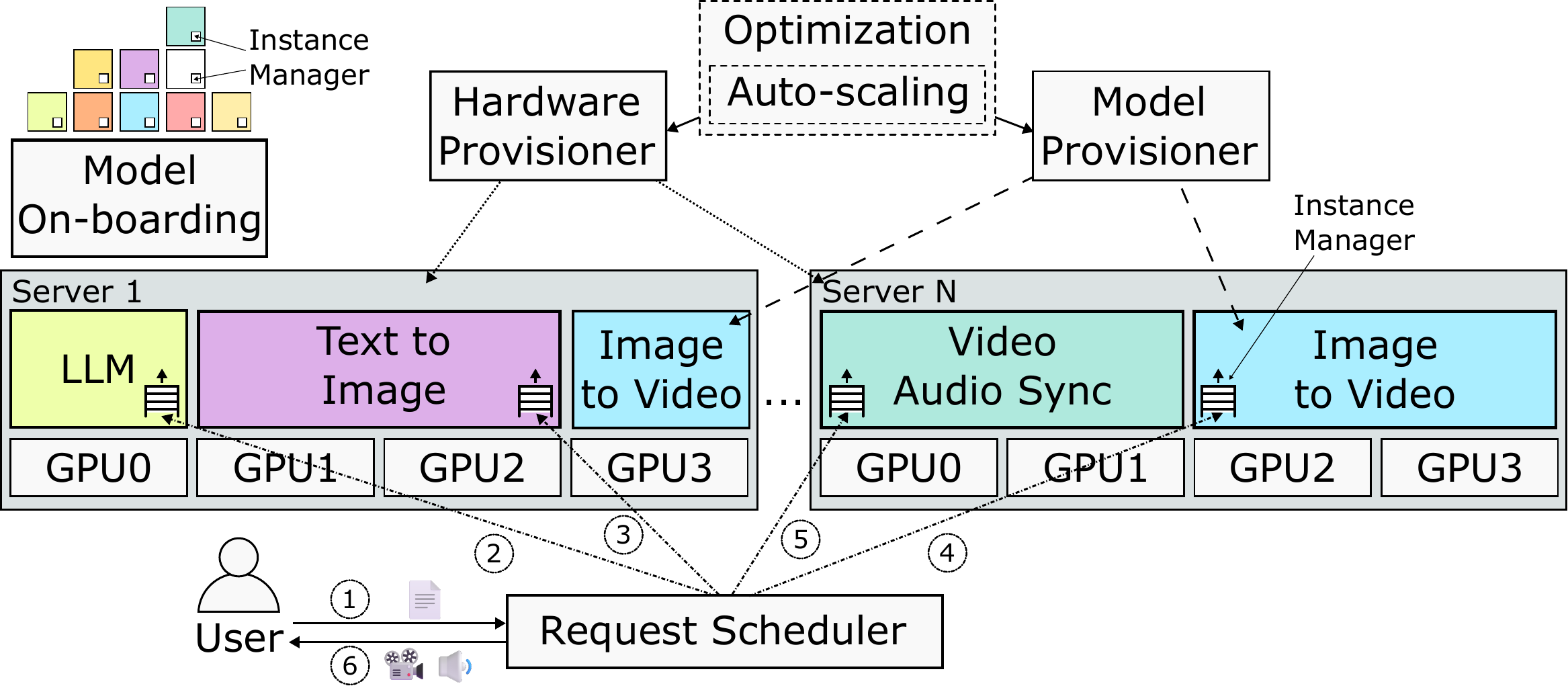}
    \vspace{-20pt}
    \caption{Architecture overview for \system{} showing an example request scheduling across four model instances.}
    \label{fig:sys_architecture}
    \vspace{-10pt}
\end{figure}

\Cref{fig:sys_architecture} shows the high-level architecture for \system{} which includes:
(1) on-boarding multi-modal generation models,
(2) provisioning resources (\ie{}, number of servers and GPUs) and models (\ie{}, two replicas of Flux and five Fantasy Talking replicas) in the cluster,
(3) scheduling and orchestrating multi-modal generation requests across model instances to fulfill latency requirements (\eg{}, real time),
and
(4) execute requests in the local model instance.

\subsection{On-boarding multi-modal models}
New models with varying performance and cost trade-offs are released weekly~\cite{hidream,fluxkontext,flux2024,kokoro,wu2024janus} (\Cref{tab:models}).
\system{} supports rapid on-boarding:
licensing and usage restrictions are handled up front, and once packaged, models can be deployed and served with minimal effort.
To allow quick integration and on-boarding, we package each model with our \emph{instance manager} which offers a standard remote interface.

\myparagraph{Characteristics}
During on-boarding, we describe the model characteristics in a metadata file that \system{} later uses for model selection.
These characteristics include
quality (Elo ranking),
frame rate (FPS),
maximum number of frames (video length),
number of attention heads,
VAE compression ratios,
supported resolutions, and
other relevant attributes.

\myparagraph{Profiling}
We generate simple model profiles to estimate runtime and resource usage, as key parameters (\eg{}, pixel count, frame count) scale proportionally.
We benchmark a representative configuration (\eg{}, 1+16 frames, 10 steps, 640$\times{}$400 resolution) and validate it against additional test points.
We also measure peak power, energy, and temperature.
These data inform predictive models for performance, cost, and quality under different configurations.

\subsection{Provisioning hardware and models}
We frame hardware and model selection for a workload (\eg{}, a 10-minute medium-quality video podcast) as an optimization problem.
After selecting a configuration, the \emph{hardware} and \emph{model provisioners} handle setup accordingly.

\myparagraphemph{Auto-scaling}
To adapt to changing demand, \system{} runs the optimization periodically.
We start from the current configuration and add the costs for provisioning the resources and the additional time to load the models.
In this way, \system{} may scale-out to ten H200 servers during daytime and scale-in to a single A100 server at night.

\myparagraph{Optimization process}
It consists of two phases:

\myparagraphemph{Initial provisioning}
We start with a cost-efficient baseline configuration that leverages inexpensive models and lower-cost GPUs.
Each model instance (\eg{}, Flux for image generation, Hunyuan FramePack for video) is assigned a single GPU (\eg{}, A100).
A greedy algorithm simulates how user requests would be processed in this setup.
It represents requests as directed acyclic graphs (DAG) and prioritizes node assignments along the critical path to available resources.
Latency and cost estimates are derived from model profiles generated during on-boarding.

\myparagraphemph{Iterative refinement}
Next, we improve the initial configuration by systematically exploring the latency-cost trade-off space.
For each setting, we use the greedy algorithm to estimate the latency and cost.
If a solution is unfeasible (\eg{}, no image generation models), it gets discarded.
This refinement process includes:
(1) adding or removing hardware resources (including Spot);
(2) switching GPU types;
(3) switch model for a task; and
(3) adjusting the number of model instances; and
(4) modifying GPU allocation per instance (\ie{}, model parallelism).
We also add domain-specific heuristics to guide the evolution.
If the cost exceeds the budget, we switch to Spot VMs and scale-in VMs.
If latency is too high, we scale-out and try faster GPUs.

\myparagraph{Optimization objective}
We minimize cost $\times$ TTFF (\$ $\times$ seconds).
When targeting specific SLOs (\eg{}, real-time with a 10-second TTFF), we steer the evolution process toward configurations that satisfy this.
If we cannot find feasible solutions, it returns the closest solution.
During exploration, we navigate the Pareto frontier between latency and cost.
By default, the system balances both objectives, but this behavior can be tuned to prioritize one.
We also support optimizing for energy, TTFF, and other combinations (\eg{}, Energy $\times$ TTFF).

\myparagraph{Optimization extensions}
To support several advanced refinements, we extend our basic optimization framework.

\myparagraphemph{Quality}
\system{} can choose between generating video directly at high quality or producing a lower-resolution version and using a separate model for up-scaling.
For example, we can run Fantasy Talking at $640\times400$ and then upscale it to $1280\times800$ using Real-ESRGAN.

\myparagraphemph{Disaggregation}
We can disaggregate compute-intensive model components such as DiT and VAE into separate components.
FramePack DiT streams latent outputs to the VAE for decoding, which are then passed to the audio sync model (\eg{}, \Cref{fig:scheduler}).
This enables pipelined execution, independent scaling, and fine-grained resource allocation.
After disaggregation, each component is represented as a distinct DAG node used by the provisioning and optimization algorithms.

\myparagraphemph{Spot}
These instances offer significant cost savings (\eg{}, up to 50\%), naturally biasing the optimization toward their use.
However, they are subject to eviction.
We proportionally increase the number of allocated resources to the eviction risk.
This ensures that the cost reflects this risk, striking a balance between affordability and reliability.

\myparagraphemph{Multi-region}
Since GPU availability varies across regions over time, \system{} continuously monitors and aggregates this availability.
We consider both inter-region latency and bandwidth when calculating generation latency and cost.
For example, small image transfers can tolerate inter-region latency, but components like DiT and VAE should remain co-located within the same region for performance reasons.

\subsection{Scheduling requests across model instances}
\label{sec:request_sched}

Once hardware and models are provisioned, the \emph{request scheduler} orchestrates execution using a live, iterative version of our greedy algorithm informed by the request DAG.

\begin{figure}
    \centering
    \includegraphics[width=\linewidth]{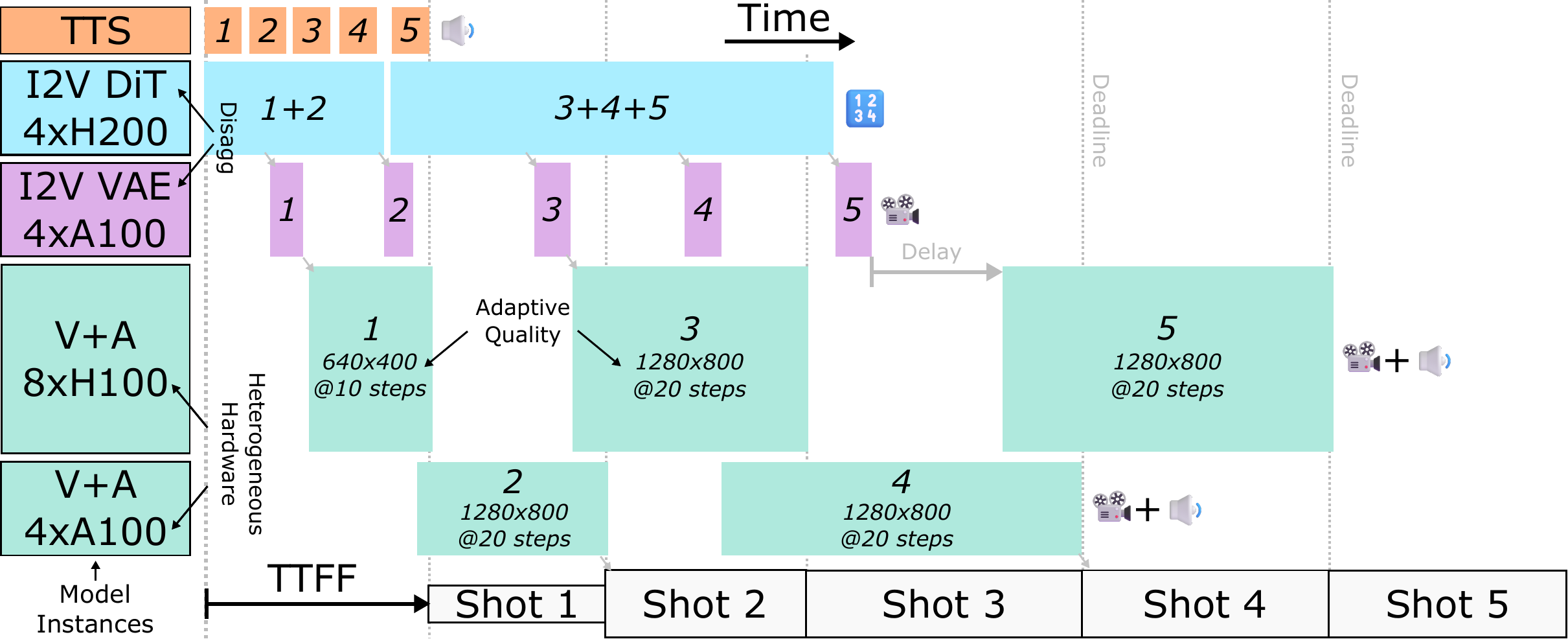}
    \caption{Deadline-aware scheduling with disaggregated components and adaptive quality on A100, H100, and H200.
    }
    \label{fig:scheduler}
\end{figure}

\myparagraph{Multiple requests}
Following the YARN philosophy~\cite{vavilapalli2013yarn}, each user request is managed by a \emph{request scheduler}.
To coordinate multiple requests, model instances maintain local queues that prioritize tasks by deadline.
For example, the image generation model may process an early scene from a new request before a later scene from an earlier request if it has a tighter deadline.
The \emph{scheduler} monitors these queues and re-schedules as resources become available.

\myparagraph{DAG generation}
Most of the DAG is generated at runtime.
For example, we do not know the number of scenes and shots until the screenplay is generated.
We start with a sketch of the DAG (\eg{}, 10-minute video with 30 seconds per shot) and as stages are generated, we update the DAG.

\myparagraph{Deadlines}
The \emph{request scheduler} computes deadlines for each DAG node based on the SLO and expected runtimes.
For example, a real-time video podcast with a TTFF of 5 seconds and 10-minute duration sets the final node’s deadline at $t_{now}+605$, with dependent nodes scheduled recursively.

Execution starts with dependency-free nodes, prioritized by deadline.
As nodes finish, their dependents are triggered.
For example, screenplay generation precedes audio and image generation, with images prioritized due to tighter deadlines.
The \emph{request scheduler} also favors earlier scenes and shots (\Cref{fig:scheduler}).
Deadlines are attached to each submission for fine-grained, instance-level scheduling.

\myparagraph{Instance selection}
Each DAG node is assigned to the model instance with the shortest expected runtime (\eg{}, 8$\times$H200 \vs{} 4$\times$A100 in \Cref{fig:scheduler}).
Lightweight tasks (\eg{}, prompt filtering, safety checks) are typically routed to cheaper GPUs or CPUs, while early-stage tasks may use higher-end instances.
If no instance is available, the \emph{request scheduler} queues the request and reschedules as resources free up.

\myparagraph{Adaptive quality}
Scheduling begins with the target quality (\eg{}, $1280\times{}800$, 20 de-noising steps) and degrades incrementally (\eg{}, $640\times{}400$, 10 steps in \Cref{fig:scheduler}) if deadlines are at risk of not being met.
Initial and intermediate stages may run at lower quality, while final outputs are rendered at higher resolution.
If not enough, we switch to static content.

\myparagraph{Model constraints}
Our greedy algorithm accounts for model-specific limits (\eg{}, maximum generation length).
For example, Fantasy Talking supports up to 3.5 seconds of video/audio at 23 FPS before drifting.
To handle longer shots, we first generate a full clip (\eg{}, 30 seconds) with FramePack at medium quality, segment it at speech pauses, and re-sync audio using Fantasy Talking.
Resolution constraints are also considered to ensure suitable aspect ratios.

\myparagraph{Evictions and failures}
Spot resources typically provide a 30-second eviction notice~\cite{yi2010reducing}.
After receiving this notice, we stop sending new requests to instances on affected resources.
For other failures, we monitor the model instances liveness and avoid sending requests to unresponsive ones.
Requests running on failed resources are resubmitted.

\myparagraph{Caching}
\system{} supports reusing intermediate results across requests to improve efficiency.
Common assets (\eg{}, static backgrounds, text/image embeddings, or previously generated segments) are cached and reused.
This can be extended to support diffusion-aware caching methods (\eg{}, AdaCache~\cite{kahatapitiya2024adacache}, NIRVANA~\cite{agarwal2024nirvana}, MoDM~\cite{xia2025modm}).

\subsection{Executing requests in a model instance}
Each model is deployed with a local \emph{instance manager}, which exposes a standardized REST interface and triggers the local model inference.
The manager also handles request \emph{batching} and adjusts GPU \emph{frequencies} to optimize resource usage.

\myparagraph{Batching}
Incoming requests are placed in a \emph{local queue} for \emph{batching} and fine-grained scheduling~\cite{yu2022orca}, with deadlines guiding reordering.
\system{} aggregates requests to boost throughput, but gains are limited:
compute-heavy stages like DiT and VAE already saturate GPUs, leaving little room for parallelism.
Lighter components (\eg{}, text/image encoders) are batched more often, though with modest benefits.

\myparagraph{Frequency management}
To align with datacenter infrastructure constraints, the \emph{instance manager} can reduce GPU frequency and introduce slight delays to reduce:
(1) peak power~\cite{stojkovic2025tapas,patel2023polcapoweroversubscriptionllm},
(2) energy~\cite{stojkovic2025dynamollm}, and
(3) thermal load~\cite{stojkovic2025tapas}.

\subsection{Implementation}
\label{sec:impl}
We build \system{} on top of a Kubernetes (K8s) cluster~\cite{burns2016borg}:
a widely adopted cluster manager that enables modular deployment, auto-scaling, service discovery, and fault tolerance.
All controllers are implemented in Python, and our implementation and deployments scripts will be open-sourced.
\system{} extends Kubernetes with a hardware and model provisioner, an instance manager, and an executor that defines, schedules, and orchestrates multi-modal real-time generation workflows.

\myparagraphemph{Alternatives}
\system{} can also be implemented on platforms such as Ray~\cite{moritz2018ray}, YARN~\cite{vavilapalli2013yarn}, for Knative~\cite{kaviani2019knative}.
Their default configuration use FIFO schedulers without deadlines.
Our scheduler adds deadline-awareness atop mechanisms similar to those in Ray and YARN (\eg{}, local queues, task stealing).
For workflow specification, one could rely on tools such as ComfyUI~\cite{comfyui}, though these still require manually disaggregating model components.
We choose Kubernetes for its simplicity and flexibility, while still retaining the same fundamental execution model as these other systems.

\myparagraph{Model on-boarding}
We package each model in \Cref{tab:models} as a Docker~\cite{merkel2014docker} container, based on an NVIDIA image with GPU drivers and runtime tools~\cite{nvidia_docker}.
Each container embeds our \emph{instance manager}, which standardizes the interface for executing inference requests.
We adapt existing inference code (typically from Hugging Face) to this interface and bundle it with the model weights.
Models using Diffusers~\cite{diffusers} can be on-boarded in <30 minutes;
others may take 1--2 hours.

All models run on PyTorch~\cite{pytorch}, with optional optimizations such as quantization~\cite{rokh2023quantization} and TeaCache~\cite{liu2024teacache}.
For LLMs (\eg{}, Llama, Gemma), we use the vLLM 0.9.1~\cite{vllm} Docker image with OpenAI APIs~\cite{openai_api}.
Metadata for on-boarded models is maintained in a centralized JSON file.

\myparagraphemph{Parallelism}
Many diffusion models include native support for multi-GPU inference (\eg{}, Wan~\cite{wang2025wan}).
For those that do not, we use USP~\cite{fang2024unified} from xDiT~\cite{fang2024xdit}.
We have enabled parallelism for four models (\eg{}, Fantasy Talking, Hunyuan FramePack), each requiring under two hours of work.
The xfuser~\cite{fang2024xdit} repository provides examples, and this process could be streamlined with LLM-based coding agents.
This is similar to the diffusion optimizations added to SGLang~\cite{sglangdifussion}.

\myparagraphemph{Profiling}
We use scikit-learn~\cite{scikit-learn} to fit linear models.
Our runtime and cost profiles are over 99.9\% accurate.

\myparagraphemph{Quality}
When on-boarding the model, \system{} uses the Elo rankings from public leaderboards~\cite{image_leaderboard,video_leaderboard}.

\myparagraph{Hardware provisioner}
We use VMs as the underlying resource pool for the K8s cluster.
The \emph{hardware provisioner} interacts with the cloud provider (\eg{}, EC2~\cite{ec2}, Azure~\cite{azure}) via standard APIs to add/remove VMs to the cluster.
It supports heterogeneous hardware (\eg{}, V100, A100, H100, H200) and Spot VMs, all interconnected within a virtual network.
Where applicable, we configure InfiniBand~\cite{infiniband} and NCCL~\cite{nccl}, and set up network peering to enable cross-region.
We expose all these features (GPU, region, evictable) as K8s node labels to be used for pod scheduling via affinity rules.

\myparagraph{Model provisioner}
The provisioning optimization completes in $\sim$90 ms and outputs the model instances to deploy (\eg{}, two FantasyTalking instances on 8$\times$H100s and 2$\times$A100s).
The provisioner uses K8s interfaces to launch these instances on the designated hardware.
This process may take several minutes, as it involves pulling Docker images, loading model weights, and performing warm-up runs.
To reduce startup latency, we cache frequently used images in each region.
Model-to-hardware mapping is managed via K8s pod affinity rules.
We also support partial-GPU deployments, allowing lightweight models (\eg{}, Kokoro) to share a GPU using MPS~\cite{nvidia_mps} and MIG~\cite{nvidia_mig}.

\myparagraph{Instance manager}
It exposes a simple HTTP REST interface using Quart~\cite{quart}.
It implements a deadline-aware request queue for local batching, enabling global coordination among \emph{request schedulers}.
This manager also sets GPU frequencies using NVIDIA interfaces.

\myparagraph{\app{}}
We implement podcast video generation as a K8s service: \app{}.
It exposes a REST API~\cite{quart} for users to submit requests (\eg{}, ``generate a 10-minute video for this paper at medium quality'').
It builds the workflow DAG with the stages, scenes, and shots (\eg{}, audio for shot 3 in scene 2).
It starts with the screenplay node, and as the LLM generates scenes, it adds nodes to the DAG.
It first creates a sketch DAG (\eg{}, 5-minute video with 45-seconds shots) with rough deadlines, later refining them as actual nodes are generated (\eg{}, scene 2 from 0:37--1:12 must complete by 19:39 UTC).

\app{} uses the \emph{request scheduler} library to discover all deployed instance models in the K8s cluster, along with their hardware resources and locations.
The \emph{scheduler} then dispatches DAG nodes to the appropriate model instances.
If a shot cannot meet its deadline, its quality is reduced.
Finally, \app{} stitches the outputs using FFmpeg~\cite{ffmpeg} and streams the result back to the user.

\myparagraph{Other applications}
In addition to \app{}, we implement the other application workflows in
\Cref{tab:other_workflows}:
StreamShort,
StreamMovie,
StreamAnimated,
StreamLecture,
StreamPersona,
StreamDub,
StreamEdit,
and StreamChat.
We mainly modify the inputs for the LLM (content and prompt with few-shot examples) and the DAG generation.
We use the same components and libraries for the rest.

\section{Evaluation}
\label{sec:eval}

\myparagraph{Methodology}
We include a \emph{naive}, non-optimized baseline that represents an out-of-the-box implementation.
This baseline statically partitions models across the available GPUs based on their maximum effective parallelism.
For example, in a deployment with four servers, each equipped with 8 A100 GPUs, it assigns GPUs to models in proportion to their runtime, subject to each model’s parallelism limits (\eg{}, Flux is capped at 8 GPUs).
Models that cannot be parallelized (\eg{}, YOLO) are pinned to a single GPU.
The baseline uses on-demand (non-spot) A100 GPUs in a single region, does not disaggregate models, and executes each request using the highest-quality (without upscaling).
We also compare to other baselines that port existing optimizations for LLMs to multi-modal generation workflows.

\myparagraphemph{Hardware}
We run our main experiments in Azure VMs using Azure Kubernetes Service (AKS)~\cite{aks_microsoft_2025}.
We evaluate a small-scale setup consisting of a single server with 8\texttimes{}A100 GPUs, as well as multiple configurations combining A100, H100, H200, and GB200 servers.
These configurations range from 8 to 40 servers and include up to 320 GPUs, for a total of 10 configurations.
We use these representative data points to validate our latency and cost estimators, and then use the validated model to simulate additional configurations.

\subsection{Model provisioning}
We first evaluate \system{} provisioning \app{} with
(1) a low-cost setup with a single A100 server with 8 GPUs and
(2) a cost-efficient setup with a maximum of 32$\times$8$\times$A100 and 8$\times$8$\times$H200 servers (320 GPUs).
\Cref{tab:gpu-single100-allocation} shows the GPUs allocated for each model.
For the A100 setup, lightweight models (Kokoro and YOLO) share a single GPU, while the most compute-intensive stage (\ie{}, Fantasy Talking) is allocated two GPUs.
In contrast, the \emph{Naive} baseline assigns one GPU per model without FramePack disaggregation, increasing TTFF from 5 hours to over 8 hours.

The cost-efficient setup uses 12 Fantasy Talking instances (96 A100 and 50 H200) to meet latency requirements.
The \emph{Naive} approach provisions 200 H100 GPUs but still suffers from 9.7\texttimes{} higher end-to-end latency.
This setup has 4 Fantasy Talking instances with 4$\times$8$\times$H200s (96 GPUs), which is substantially less effective due to limited pipelining.
The \emph{Naive} baseline consumes the same number of GPUs but increases the TTFF from 20 seconds to over 4 minutes.

\begin{table}[t]
    \centering
    \caption{
    Provisioning and generation time (in seconds) for \app{} generating a 10-minute podcast video.
    The video consists of 43 shots at 1280$\times$800, using 20 diffusion steps.
    }
    \label{tab:gpu-single100-allocation}
\footnotesize
\begin{tabular}{l crr crr}
    \toprule
                    & \multicolumn{3}{c}{Low-cost}     & \multicolumn{3}{c}{Cost-efficient} \\
                    & \multicolumn{3}{c}{8$\times$A100}& \multicolumn{3}{c}{32$\times$8$\times$A100 + 8$\times$8$\times$H200} \\
    Model           & \#GPUs &   TTFF & Time           & \#GPUs & TTFF & Time\\
    \midrule
    \app{}          & -    &   1.2 &     3.5           &    -  &  1.2 &   3.5 \\
    Gemma           & 1    &   6.6 &    31.8           &  8+0  &  1.5 &   9.7 \\
    Flux            & 1    &   9.8 &     9.8           & 16+0  &  \cellcolor{green!20}0.9 &   1.0 \\
    YOLO            & \cellcolor{green!20}0.5  &   \cellcolor{green!20}0.1 &     \cellcolor{green!20}0.6  & \cellcolor{green!20}0.5+0  &  \cellcolor{green!20}0.1 & \cellcolor{green!20}0.1 \\
    Kokoro          & \cellcolor{green!20}0.5  &   \cellcolor{green!20}0.6 &    25.8           & \cellcolor{green!20}0.5+0  &  \cellcolor{green!20}0.6 &  25.8 \\
    FramePack       & 1    &   8.6 &  1486.2           & 41+8    &  \cellcolor{green!20}0.9 &  27.1 \\
    FramePack VAE   & 1    &   2.0 &   343.0           & 20+4    &  \cellcolor{green!20}0.9 &  52.4 \\
    Fantasy Talking & \cellcolor{red!20}2 & \cellcolor{red!20}78.8 & \cellcolor{red!20}13588.7 & \cellcolor{red!20}96+50 & \cellcolor{red!20}14.6 & \cellcolor{red!20}131.1 \\
    Real-ESRGAN     & 1    &  15.4 &  2663.4           & 74+2   &  2.0 & 34.4  \\
    \bottomrule
\end{tabular}
\end{table}

Based on the model provisioning, \app{} follows the workflow in \Cref{fig:video_generation}:
(1) Gemma~\cite{gemma3} generates the screenplay, triggering generation of the first shot;
(2) Kokoro~\cite{kokoro} synthesizes the speech audio;
(3) Flux~\cite{flux2024} generates a high-quality base image;
(4) YOLO~\cite{yolo11} produces zoomed-in image crops;
(5) Hunyuan+FramePack~\cite{kong2024hunyuanvideo,zhang2025framepack} uses these images to generate a sketch video in latent space;
(6) in parallel, Hunyuan+FramePack VAE decodes the sketch video from the latent representation;
(7) Fantasy Talking~\cite{wang2025fantasytalking} (based on Wan~\cite{wang2025wan}) combines video and audio at medium quality as they are produced;
(8) Real-ESRGAN~\cite{wang2021realesrgan} upscales the output to high quality; and
(9) \app{} stitches the shots together and streams the final video.

\myparagraphemph{Multiple regions}
Constraining our deployment to a single region, limits the availability of hardware and without it, we cannot combine H200 and A100 hardware.
H200 is placed in East Us and A100 in West US.
\system{} accounts for the bandwidth and latency when provisioning the models.
This is similar to what Helix~\cite{mei2025helix} does for LLMs.
In addition, using multiple regions improves tolerance to Spot VM evictions by reducing correlation in eviction events.

\subsection{Serving a multi-modal generation request}
\label{sec:eval_single}
\Cref{tab:gpu-single100-allocation} shows the TTFF and total time to generate a 10-minute video podcast about this paper at $1280\times800$ and 23 FPS using our 8$\times$A100 and the mixed A100+H200 setups.
On 8$\times$A100 GPUs, the first frame is ready in 123 seconds, but the final frame is not generated until 3.8 hours later.
For uninterrupted real-time streaming, TTFF is 3.7 hours.
The 256$\times$A100+64$\times$H200 setup reduces the TTFF to <22 seconds.
The remaining frames are generated within 10 minutes, enabling real-time streaming right after the initial TTFF.

\begin{figure}[t!]
    \centering
    \includegraphics[width=\linewidth]{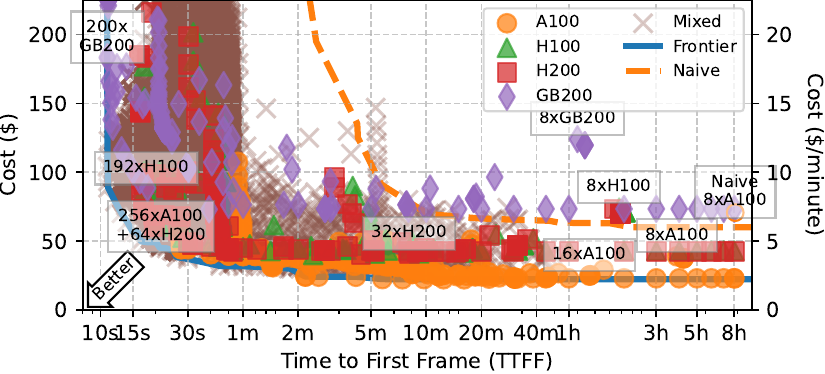}
    \caption{TTFF \vs{} cost generating a 10-minute high-quality video podcast using various hardware configurations.
    The cheapest and fastest setup lies in the bottom-left corner.}
    \label{fig:eval_latency_vs_cost}
\end{figure}

\myparagraph{Latency \vs{} cost}
\Cref{fig:eval_latency_vs_cost} shows the TTFF and cost to generate a 10-minute video podcast at high quality changing the provisioned hardware.
The 8$\times$A100 GPUs setup is <\$25 but incurs high latency.
For reference, running naively in 8$\times$A100 without disaggregation, upscaling, and deadline-aware scheduling is 2.2\texttimes{} slower and >\$50 more expensive.
The A100+H200 configuration is slightly more expensive at around \$45 but achieves a TTFF under 22 seconds.
We evaluated both setups on a real Azure Kubernetes cluster.

64$\times$A100 achieves a 2-minute TTFF at around \$25.
In contrast, 8$\times$A100 is nearly 2\texttimes{} more expensive than 16$\times$A100 due to longer execution time.
Mixing A100 and H200 reduces TTFF to under 1 minute while keeping costs below \$50.
Overall, A100 offers the best cost efficiency, while mixed A100+H100 configurations provide the best latency--cost trade-off.
H200 configurations deliver marginal performance gains over H100, making them less cost-effective.
GB200 further reduces TTFF, but its significantly higher hourly cost makes it uncompetitive with H100 except when targeting very low latencies (\ie{}, under 15 seconds).
\system{} provisions a small number of expensive GPUs (\eg{}, GB200) to reduce TTFF and allocates the remaining capacity to cheaper GPUs (\eg{}, A100).
\system{} selects the cost-efficient option with 256$\times$A100 and 64$\times$H200.

\myparagraphemph{Relaxed SLOs}
When serving a request without a real-time deadline, \system{} navigates this latency space and chooses the cheapest configuration.
For example, when the video does not need to be generated within 2 hours, it uses 2 servers with 8$\times$A100 GPUs, reducing the cost to under \$25.
Costs remain low until the TTFF deadline tightens to 2 minutes.

\begin{figure}[t!]
    \centering
    \includegraphics[width=\linewidth]{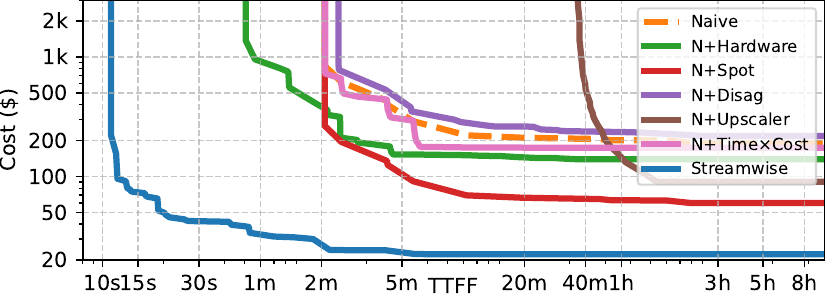}
    \caption{
    Contribution of each technique used by \system{} at high quality, including the \emph{Naive} baseline.
    }
    \label{fig:ablation}
\end{figure}

\myparagraph{Ablation studies}
\Cref{fig:ablation} shows the contribution of each \system{} technique relative to the \emph{Naive} baseline.
\emph{Hardware} reduces both cost and latency.
\emph{Disaggregation} further reduces cost, while \emph{Spot} lowers it with similar latency.
Optimizing for \emph{Time\texttimes{}Cost} significantly reduces cost with minimal latency impact.
\emph{Upscaler} substantially reduces TTFF but has little effect on cost.
Combining all techniques in \system{} achieves the best cost--latency trade-off.

\begin{figure}[t!]
    \centering
    \includegraphics[width=\linewidth]{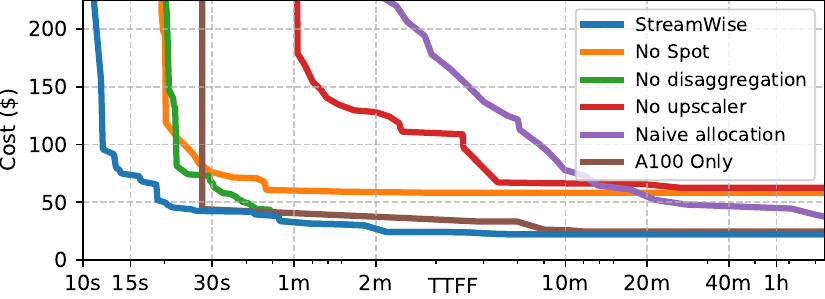}
    \caption{
    Penalty of disabling each technique in \system{} when generating a 10-minute podcast at high quality.
    }
    \label{fig:ablation_minus}
\end{figure}

\Cref{fig:ablation_minus} shows the penalty of omitting individual techniques from \system{}.
Without \emph{hardware} (using \emph{A100-only}), TTFF cannot go lower than 30 seconds.
Without \emph{spot}, costs increase substantially, shifting the frontier upward.
Without \emph{disaggregation}, the system cannot leverage pipelining, resulting in higher TTFF.
Without \emph{upscaler}, both cost and latency increase significantly.
Replacing the greedy heuristic with the \emph{naive allocator} also substantially increases cost.

Overall, these results show that no single technique is sufficient to enable efficient real-time generation.

\begin{figure}
    \centering
    \includegraphics[width=\linewidth]{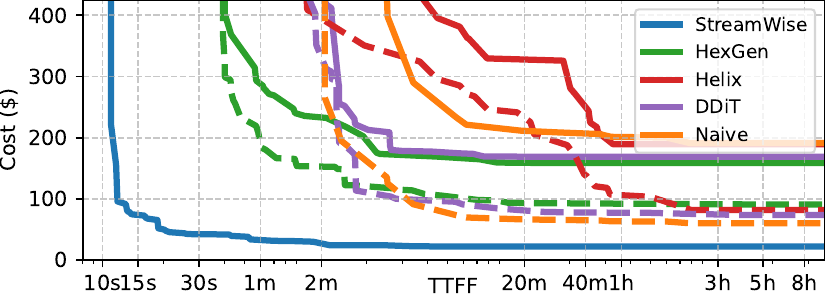}
    \caption{Applicability of systems for efficient LLM serving.
    Dashed lines use Spot VMs to reduce cost.}
    \label{fig:eval_llm}
\end{figure}

\myparagraph{Porting LLM serving optimizations}
Applying LLM prefill/decode disaggregation~\cite{patel2023splitwise,zhong2024distserve} does not improve TTFF or cost for multi-modal workflows, as LLMs account for less than 1\% of total execution time.
As a result, optimizations that target LLM serving in isolation have negligible end-to-end impact.
Nevertheless, we adapt two representative systems (HexGen~\cite{jiang2023hexgen} and Helix~\cite{mei2025helix}) and redesign them for our setting.
We further extend both to optionally leverage Spot VMs, making them more competitive with \system{}.
\Cref{fig:eval_llm} compares their TTFF and cost against \system{}.

HexGen~\cite{jiang2023hexgen} partitions LLMs using tensor and pipeline parallelism and uses a genetic algorithm to search over placement and parallelization strategies across heterogeneous GPUs, targeting high throughput.
We generalize it to treat each component in the multi-modal pipeline as a partitionable unit and use its search to determine placement, allocation, and parallelism.
Despite this adaptation, Spot HexGen incurs over 3\texttimes{} higher cost and achieves a TTFF that is 5\texttimes{} slower than \system{}.
This is because HexGen optimizes per-model throughput rather than minimizing end-to-end critical-path latency across heterogeneous stages.

Helix~\cite{mei2025helix} formulates LLM serving as a max-flow problem and uses MILP to compute optimized placement and parallelization strategies.
We extend its formulation to jointly optimize all components in the multi-modal pipeline.
However, Spot Helix performs even worse in TTFF, as it optimizes each model independently within a global budget, without balancing resources across pipeline stages.
As a result, it performs even worse than \emph{naive} as it over-provisions some models while under-provisioning others, leading to stage imbalance and elongated critical paths.
This per-model optimization fails to capture cross-stage dependencies, degrading end-to-end latency.

\myparagraph{Disaggregation}
Recent systems such as DDiT~\cite{huang2025ddit} and StreamDiT~\cite{kodaira2025streamdit} advocate disaggregating diffusion-based models to improve utilization and scalability.
\Cref{fig:eval_llm} includes this DiT-disaggregated design applied to the podcast generation workflow.
However, separating the DiT and VAE components alone is insufficient to achieve efficient real-time performance. 
While disaggregation increases resource flexibility, it does not reduce the critical-path latency. %

The cost impact also depends strongly on deployment scale.
At small scales, disaggregation increases cost because DiT and VAE must run on separate GPUs instead of being colocated on the same device.
However, at larger scales, it enables independent scaling of each component (\eg{}, 32 GPUs for DiT but only 2 for VAE), allowing capacity to better match demand and reducing cost.
Although DiT/VAE disaggregation improves elasticity at scale, it is insufficient by itself to enable efficient real-time multi-modal generation.
This observation is consistent with \Cref{fig:ablation_minus}, which shows \system{} without disaggregation.

\begin{figure}
    \centering
    \includegraphics[width=\linewidth]{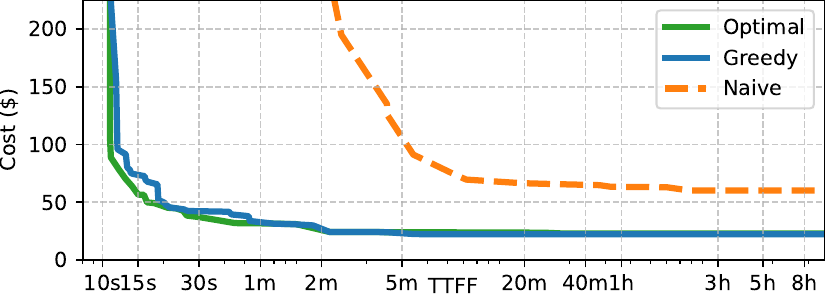}
    \caption{Quality of the greedy heuristic against the optimal allocation. Optimal takes over 100\texttimes{} longer than greedy.}
    \label{fig:eval_greedy_milp}
\end{figure}

\myparagraph{Greedy heuristic}
\Cref{fig:eval_greedy_milp} compares \system{}’s greedy allocation heuristic with an optimal formulation that uses Gurobi~\cite{gurobi} to solve the mixed-integer program for assigning multi-modal AI components to GPUs.
For relaxed latency targets (\ie{}, TTFF above one minute), the greedy heuristic matches the optimal solution.
Under stricter latency constraints, it remains within 20\% of optimal cost and converges to the same solution for the tightest TTFF targets.

The computational overheads of these algorithms however, differ dramatically.
The greedy heuristic completes in under 100\,ms, enabling online reconfiguration.
In contrast, solving the full mixed-integer program can take over 10 minutes, making it impractical for real-time control.
When higher accuracy is needed, cached optimal solutions can warm-start the greedy heuristic.

\begin{figure}[t!]
    \centering
    \includegraphics[width=\linewidth]{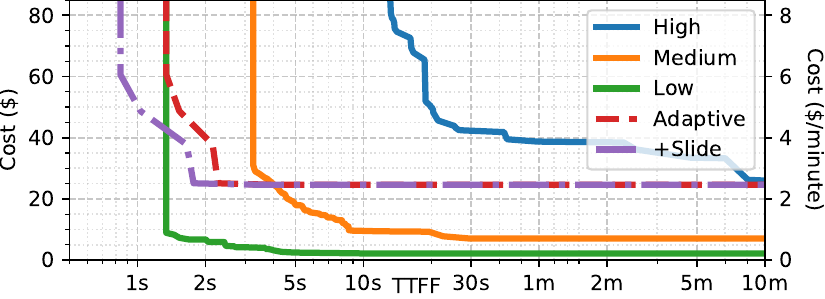}
    \caption{
    TTFF \vs{} cost Pareto frontiers with high, medium, and low output qualities.
    Includes our adaptive quality policy.
    }
    \label{fig:eval_quality}
\end{figure}

\myparagraph{Adaptive quality}
\Cref{fig:eval_quality} shows the impact of generating content at three qualities:
(1) high quality with 20 denoising steps and $1280\times{}800$ (same as \Cref{fig:eval_latency_vs_cost}),
(2) medium with 10 steps at $640\times{}400$, and
(4) low with 5 steps at $320\times{}200$.
With low quality, we achieve a TTFF <3 seconds for <\$0.5/minute.

\myparagraphemph{Adaptive}
\Cref{fig:eval_quality} shows how adjusting quality during content generation reduces latency.
Streaming begins at low quality with a TTFF under 3 seconds, switches to medium quality after 10--20 seconds, and reaches high quality within 45 seconds. 
This approach costs under \$50 while delivering high quality for over 90\% of the video.

\myparagraphemph{Non-generated content}
We further cut TTFF under 1 second by cleverly starting with a 500 ms title slide (\eg, with the paper title and the front page), accompanied by a host voice-over introducing the video.

\begin{figure}
    \centering
    \includegraphics[width=\linewidth]{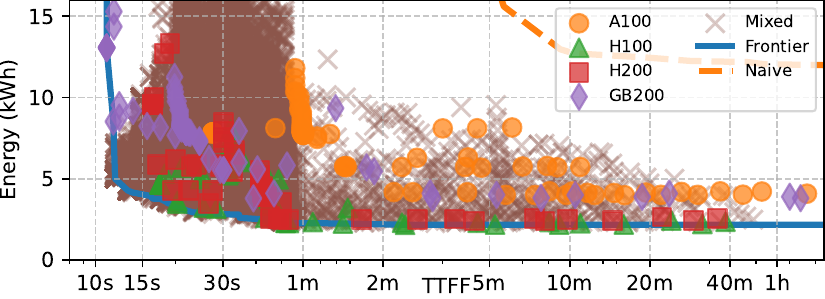}
    \caption{
    Energy efficiency for video podcast generation.
    }
    \label{fig:eval_energy}
\end{figure}

\myparagraph{Energy efficiency}
\Cref{fig:eval_energy} shows the energy efficiency to generate a 10-minute podcast.
In this case, A100 GPUs by themselves are not as efficient and consume 2\texttimes{} the energy compared to H100.
The energy efficiency of GB200 is similar to the one for A100.
Combining H100 with a few A100 consumes $\sim$2 kWh for a TTFF of less than 1-minute; \system{} selects this point.
In contrast, the \emph{Naive} baseline consumes over 10 kWh for the most energy efficient option and over 50 kWh for the fastest option with a TTFF of over 2 minutes.

\begin{figure}
    \centering
    \includegraphics[width=\linewidth]{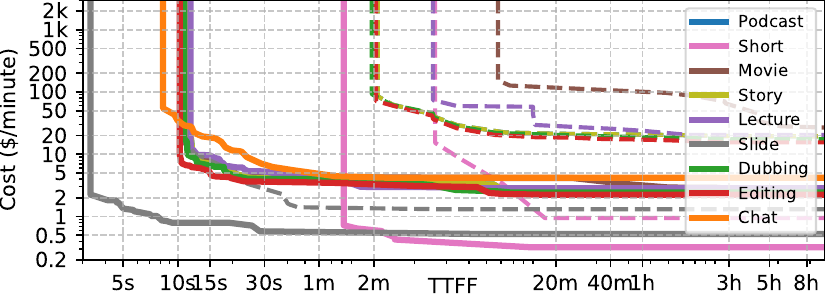}
    \caption{
    Sensitivity to the multi-modal workflow.
    Dashed lines denote \emph{Naive}, while solid lines represent \system{}.
    }
    \label{fig:eval_workflows}
\end{figure}

\myparagraph{Other workflows}
\Cref{fig:eval_workflows} evaluates all the workflows from
\Cref{tab:other_workflows}.
\system{} consistently outperforms the \emph{Naive} policy with 10.4\texttimes{} lower latency and 17.5\texttimes{} cost savings.
The most cost efficient application is \emph{Slide} because of its much smaller resolution and the most expensive one is \emph{Chat} because of its interactivity requirements.
This demonstrates that applications composed of generative, multi-modal models operating under latency and quality constraints benefit from StreamWise's abstractions and scheduling mechanisms.

\myparagraph{Monolithic model}
As no end-to-end podcast generation model exists (\Cref{sec:monolith}), we emulate one using a single diffusion model (\ie{}, Fantasy Talking) to generate a 10-minute video matching \app{}.
While \system{} still applies deadline-aware scheduling, disaggregation, and spot instances, it forgoes modular benefits such as model specialization, heterogeneous hardware across stages, and selective upscaling.
This restricted variant of \system{} closely mirrors StreamDiffusionV2~\cite{feng2025streamdiffusionv2}.

Even under this constrained setup, \system{} reduces cost by 3.5\texttimes{} relative to the monolithic baseline.
For the fully modular \app{} workflow, \system{} achieves an additional 2.7\texttimes{} cost reduction and a 4.3\texttimes{} latency reduction compared to the monolith, quantifying the performance and efficiency gains enabled by workflow-level modularity.

\subsection{Serving multiple requests}
\label{sec:eval_multi}

\begin{figure}[t!]
    \centering
    \includegraphics[width=\linewidth]{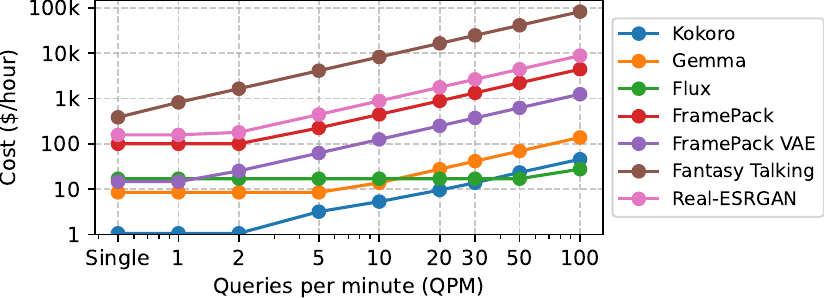}
    \vspace{-22pt}
    \caption{Model-wise cost breakdown serving video generation requests using our cost-efficient adaptive quality policy.}
    \label{fig:eval_cost_qpm}
    \vspace{-10pt}
\end{figure}

\Cref{fig:eval_cost_qpm} shows hourly costs as queries per minute (QPM) increase, based on the number of required model replicas.
It starts from the \emph{single} request from \Cref{fig:eval_quality} with 256$\times{}$A100 and 64$\times{}$H200 (the most cost-efficient).
At low rates (\eg{}, 1 QPM), minimal replicas are enough, keeping costs under \$1.2K/hour.
As load grows, more in-flight requests require additional replicas to avoid queuing and latency.

Lightweight models (\eg{}, Kokoro) benefit from instance and GPU sharing, improving cost efficiency.
Scaling Kokoro from 1 to 100 QPM increases cost by only 43\texttimes{}.
Flux achieves high throughput due to fewer activations per request, allowing one instance to serve many requests.
Fantasy Talking requires dedicated GPU replicas per request to meet real-time SLOs, limiting multiplexing and driving costs up.

The \emph{Naive} approach requires 5.6\texttimes{} higher cost to maintain the same throughput and latency.
\system{} dynamically provisions resources to efficiently handle concurrent multi-modal generation. 
Sharing lightweight models minimizes overhead, while heavier models remain the primary scalability bottleneck.

\myparagraph{Relaxed SLOs}
We extend the workload with heterogeneous latency requirements: one-third real-time, one-third with 50\% relaxed SLOs, and one-third with no SLO (\ie{}, batch).
Our deadline-aware scheduler exploits this slack to prioritize real-time requests and reduce replica count, yielding an additional 37.9\% cost reduction over the homogeneous real-time–SLO setting.
In contrast, \emph{Naive} cannot leverage this heterogeneity and continues to overprovision.
Overall, we reduce cost by 9.1\texttimes{} relative to the baseline.

\section{Related Work}
\label{sec:related}

\myparagraph{Workflow and DAG scheduling}
General-purpose cluster schedulers like Tez~\cite{saha2015apache}, Ray~\cite{moritz2018ray}, Spark~\cite{zaharia2012spark}, YARN~\cite{vavilapalli2013yarn}, Borg~\cite{burns2016borg}, and others~\cite{schwarzkopf2013omega,burns2016borg,hindman2011mesos,isard2007dryad}, established the foundation for distributed scheduling and resource management.
However, they require adaptation to meet the demands of multi-modal generation (\eg{}, \system{} builds on K8s).

\myparagraph{Inference-as-a-Service}
Early systems for ML inference, such as FastFlow~\cite{zhang2025fast}, INFaaS~\cite{romero2021infaas} target single-modality or coarse-grained batch workflows, often assuming simpler DAGs.
In contrast, modern multi-modal generation pipelines introduce fine-grained DAGs with tightly coupled components (\eg{}, DiT, VAE, upscaling, TTS), each with unique latency and resource requirements.
Llama~\cite{romero2021llama}, ServerlessLLM~\cite{fu2024serverlessllm}, and VIVA~\cite{kang2022viva} optimize cost, latency, and resource utilization.
However, they assume less dynamic operator graphs and limited interactivity.

\myparagraph{LLM serving}
Prior work focuses on optimizing the execution of LLMs which have two well-defined stages (\ie{}, prefill and decode) and KV-caches~\cite{agrawal2024taming,patel2023splitwise,zhong2024distserve,mei2025helix,jiang2023hexgen}.
SplitWise~\cite{patel2023splitwise} and DistServe~\cite{zhong2024distserve} propose disaggregating these two stages and even run them on separate hardware.
Helix~\cite{mei2025helix} and HexGen~\cite{jiang2023hexgen} propose serving LLMs across heterogeneous hardware and across regions by managing fine-grain parallelism.
Medha~\cite{agrawal2024medha} performs LLM-specific optimizations like adaptive chunking and deadline-aware scheduling.
In contrast, multi-modal generation is a more complex workflow (\eg{}, text/image encoders, VAE, DiT, TTS, LLM) with more knobs (\eg{}, resolution, number of frames and steps).

\myparagraph{Multi-modal serving}
Recent work has explored the challenges of serving models with multi-modal inputs~\cite{qiu2025modserve}.
Systems such as Mora~\cite{yuan2024mora}, StreamDiT~\cite{kodaira2025streamdit}, and StreamDiffusionV2~\cite{feng2025streamdiffusionv2} begin to address compositional and streaming text-to-video generation.
However, comprehensive solutions for real-time orchestration of multi-modal models remain limited.
Commercial systems including Veo 3~\cite{veo3}, Runway~\cite{runway}, SeeDance~\cite{gao2025seedance}, and Sora~\cite{sora} demonstrate the potential of large-scale multi-modal generation, but their system architectures and scheduling strategies are largely proprietary and unknown.

\myparagraph{Diffusion optimizations}
Recent work reduces diffusion costs through inference reuse and architectural specialization.
For text-to-image, methods like Approximate Caching~\cite{agarwal2023approximate}, TeaCache~\cite{liu2024teacache}, MoDM~\cite{xia2025modm}, and DiffServe~\cite{ahmad2024diffserve} cache internal states or adjust quality, but are limited to static, single-modality settings.
For video, AdaCache~\cite{kahatapitiya2024adacache}, FasterCache~\cite{lv2024fastercache}, and Vchitect~\cite{fan2025vchitect} use latent caching and parallelism to address batch bottlenecks. 
Personalization and adaptation often use LoRA~\cite{hu2021lora} and related techniques~\cite{smith2023continual,huang2024context}.
While optimizations like FlashAttention~\cite{dao2022flashattention,dao2023flashattention2} and SAGE~\cite{zhang2024sageattention} improve stage efficiency, they do not address end-to-end scheduling or real-time multi-modal serving.

\section{Conclusions}

We introduced \system{}, a real-time multi-modal generation system for workflows like podcast video creation.
Our modular design supports adaptive scaling, parallelism, and deployment across diverse hardware.
We identify key challenges in serving these models under strict latency constraints and quantify latency-cost-quality trade-offs.
Future work can improve efficiency through model specialization and tighter model-scheduler co-design.

\bibliographystyle{plain}
\balance
\bibliography{ref}

\end{document}